\documentclass[%
reprint,
superscriptaddress,
 amsmath,amssymb,
aps,
pra,
floatfix,
]{revtex4-2}

\usepackage{graphicx}% Include figure files
\usepackage[colorlinks=true]{hyperref}% add hypertext capabilities
\usepackage{amsmath,bbold,dsfont,xcolor}
\usepackage{braket,adjustbox}
\usepackage[makeroom]{cancel}
\usepackage{siunitx,dsfont}

\newcommand{\figref}[1]{\figurename~\ref{#1}}
\newcommand{\secref}[1]{Sec.~\ref{#1}}

\renewcommand{\eqref}[1]{Eq.~(\ref{#1})}
\newcommand{\Schr}{Schr\"odinger }
\newcommand{\ip}{{I}_{\mathrm{p}}}
\newcommand{\up}{{U}_{\mathrm{p}}}

\newcommand{\pb}{\mathrm{\mathbf{p}}}

\newcommand{\kb}{\mathrm{\mathbf{k}}}
\newcommand{\rb}{\mathrm{\mathbf{r}}}
\newcommand{\Ab}{\mathrm{\mathbf{A}}}
\newcommand{\Eb}{\mathrm{\mathbf{E}}}
\newcommand{\Bb}{\mathrm{\mathbf{B}}}
\newcommand{\pbh}{\hat{\mathrm{\mathbf{p}}}}
\newcommand{\pbt}{\tilde{\mathrm{\mathbf{p}}}}
\newcommand{\rbh}{\hat{\mathbf{r}}}
\newcommand{\pbth}{\hat{\tilde{\mathrm{\mathbf{p}}}}}
\newcommand{\SO}{\mathrm{SO}}
\newcommand{\id}{\mathds{1}}
\newcommand{\Sb}{\mathbf{S}}
\newcommand{\Lb}{\mathbf{L}}

\renewcommand\Re{\mathrm{Re}}
\renewcommand\Im{\mathrm{Im}}

\newcommand{\iden}{\mathds{1}}

\DeclareSIUnit{\au}{{a.u.}}
\DeclareSIUnit{\nm}{{nm}}

\newcommand{\orcid}[1]{%
  ~\href{https://orcid.org/#1}{\includegraphics[width=8pt]{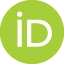}}%
  }

\newcommand{\Qprop}{Q{\textsc{prop}}}
\begin{document}

\preprint{APS/123-QED}

\title{Relativistic and Spin-Orbit Dynamics at Non-Relativistic Intensities in Strong-Field Ionization} 
%Semi-classical Spin in Strong-Field Ionization}% Force line breaks with \\
%\thanks{A footnote to the article title}%

	\author{Andrew S. Maxwell\orcid{0000-0002-6503-4661}}
	\email{andrew.maxwell@phys.au.dk}
	\affiliation{Department of Physics and Astronomy, Aarhus University, DK-8000 Aarhus C, Denmark}

\author{Lars Bojer Madsen\orcid{0000-0001-7403-2070}}
\affiliation{Department of Physics and Astronomy, Aarhus University, DK-8000 Aarhus C, Denmark}

%\collaboration{CLEO Collaboration}%\noaffiliation

\date{\today}% It is always \today, today,
             %  but any date may be explicitly specified

\begin{abstract}
Spin-orbit dynamics and relativistic corrections to the kinetic energy in strong-field dynamics, have long been ignored for near- and mid-IR fields with intensities $10^{13}$--$10^{14}$~W/cm$^2$, as the final photoelectron energies are considered too low for these effects to play a role. However, using a precise and flexible path-integral formalism, we include all correction terms from the fine-structure, Breit-Pauli Hamiltonian. This enables a treatment of spin, through coherent spin-states, which is the first model to use this approach in strong-field physics. We are able to show that the most energetically rescattered wavepackets, undergo huge momentum transfer and briefly reach relativistic velocities, which warrants relativistic kinetic energy corrections. 
We probe these effects and show that they yield notable differences for a $1600$~nm wavelength laser field on the dynamics and the photoelectron spectra. 
Furthermore, we find that the dynamical spin-orbit coupling is strongly overestimated if relativistic corrections to kinetic energy are not considered.
Finally, we derive a new condition that demonstrates that relativistic effects begin to play a role at intensities orders of magnitude lower than expected.
Our findings may have important implication for imaging processes such as laser-induced electron diffraction, which includes high-energy photoelectron recollisions.
\end{abstract}

%\keywords{Suggested keywords}%Use showkeys class option if keyword
                              %display desired
\maketitle

%\tableofcontents

\section{\label{sec:Intro}Introduction}

Electron spin is a canonical example of the departure of the atomic scale world from our macroscopic one. The discovery of spin \cite{uhlenbeck_ersetzung_1925} carried huge fundamental significance, while also leading to a vast range of observable effects, from fine structure splitting and the Zeeman \cite{zeeman_effect_1897} effect, to the Stern-Gerlach experiment and Mott scattering \cite{mott_scattering_1929}. There are a huge range of applications exploiting spin, including spin resonance imaging processes, atomic clocks, quantum sensors for magnetic fields, spin qubits, and spintronics (spin transport in solids) \cite{mott_scattering_1929}.
Despite its significance, until recently \cite{barth_spinpolarized_2013,hartung_electron_2016}, the role of spin was ignored in strong-field processes driven by ``non-relativisitic'' laser intensities. The argument given is that at these intensities, the spin will not interact with the laser field, and that the energies reached in laser driven recollisions (often taken to be $\approx3.17\up$ \cite{corkum_plasma_1993}, $\up$ is the ponderomotive energy, i.e., the cycle-averaged kinetic energy of a free electron in the laser field) are insufficient for spin-orbit coupling or other relativistic effects to play a role.

Strong-field physics deals with the interaction of intense and short laser fields with matter. Through the control of a few recollision-based processes, this has enabled the measurement and manipulation of matter on the scale of attoseconds ($10^{-18}$~s), giving birth to the field of attosecond physics \cite{krausz_attosecond_2009, salieres_study_1999, lewenstein_principles_2009, ciappina_attosecond_2017} and attosecond laser pulses \cite{paul_observation_2001,hentschel_attosecond_2001} via high-harmonic generation (HHG) \cite{corkum_plasma_1993,lewenstein_theory_1994}.
One of the primary processes involved is above-threshold ionization (ATI) \cite{agostini_freefree_1979}, which, in this context, is the strong-field (or tunnel) removal of an electron by the laser. After ionization, the continuum electron may undergo laser driven elastic scattering off the residual ion, sometimes referred to as high-order ATI (HATI) \cite{paulus_rescattering_1994,lewenstein_rings_1995}.
%The recollision picture in the gas phase follows three steps \cite{lewenstein_theory_1994,paulus_rescattering_1994} (i) strong-field (or tunnel) ionization by the laser; (ii) laser-driven propagation of the continuum electronic wavepacket; and (iii) the continuum wavepacket may be detected leading to direct-above-threshold ionization (ATI), elastically recollide with the parent ion leading to high-order above-threshold ionization (HATI), recombine with the parent ion emitting a high-energy photon in the process of high-harmonic generation, and inelastic recollide that can lead to processes such as non-sequential multiple ionization or frustrated tunnelling. HHG has gained particular notoriety for the ability to use it to generate XUV attosecond laser pulses, which along X-ray free-electron lasers (XFEL) has provided a leading role in attosecond imaging.
The processes of ATI and HATI have found use in imaging procedures, laser-induced electron diffraction (LIED) \cite{zuo_laserinduced_1996,niikura_sublasercycle_2002,amini_chapter_2020a,sanchez_molecular_2021,giovannini_new_2023} and photoelectron holography \cite{huismans_timeresolved_2011,hickstein_direct_2012,figueirademorissonfaria_it_2020} %\textbf{I understand that you cite 2 original works here for LIED and holography, and then 1 review. Should we add more including LIED work of Biegert?}
%I cited two of their works now
. In the former, the recolliding electron is used to provide diffraction images of its parent molecule. In photoelectron holography, the interference of electronic wavepackets that recollide and those that do not, is used to image the parent atom or molecule. In the case of LIED, long wavelengths, on the order of a few microns, are used to achieve high electronic recollision velocities. These hard recollisions closely probe the target, potentially leading to large spin-orbit coupling and large kinetic energies.%, meaning that the spin of the returning electron should be accounted for here. 
%However, such dynamical spin effects have not been accounted for in strong-field physics.

%Although, spin dynamics of recolliding electrons at non-relativistic laser intensities has not been studied in detail, spin in the context of strong laser fields has been investigated in other contexts. 
%TODO
%\textbf{In the following paragraph some of the works [25]-[26] are in the stabilization regime, i.e., very high frequency, with frequency higher than the ionization potential. I don't recall if the Klaiber works are for infrared light or also in the high intensity regime. Maybe we could focus in this paragraph on near-infrared and infrared wavelenghts and relativistic intensities? And then just mention very briefly that we are not concerned with the high-intensity, high-frequency stabilization regime (Gavrila, Joachian,...) - I think this could help the reader.   }
In the relativistic laser intensity regime ($I>10^{16}$~W/cm$^2$), see e.g., Refs.~\cite{reiss_complete_1990,reiss_relativistic_1990,protopapas_atomic_1997,walser_spin_2002a,klaiber_abovethreshold_2013a,klaiber_abovethreshold_2013,protopapas_atomic_1997,klaiber_spin_2014,klaiber_limits_2017,klaiber_strongfield_2017,klaiber_relativistic_2023a}, spin is considered by necessity, note we have neglected works on high intensity and high frequency (in the so-called stabilization regime), as we are focused on fields with near-infrared and infrared frequencies.
Popular ionization models such as the strong-field approximation (SFA) \cite{keldysh_ionization_1965,faisal_multiple_1973,reiss_effect_1980} were generalized to the relativistic regime \cite{reiss_complete_1990,reiss_relativistic_1990}, and many studies have considered spin related effects, see e.g., Refs.~\cite{walser_spin_2002a,klaiber_spin_2014}. However, in order to estimate the intensity at which relativistic effects become relevant, it is common to employ classical Coulomb-free trajectories \cite{protopapas_atomic_1997}, which predicts that these effects should be negligible for intensities $I<10^{16}$~W/cm$^2$. This is a good approach for the direct ATI electrons, but in the case of rescattered electrons, the Coulomb potential can allow for much higher kinetic energies to be reached. In this work, we will introduce a Coulomb adapted condition for backscattered trajectories. Some relativistic models have extended the relativistic SFA to account for the Coulomb potential, see e.g., Refs.~\cite{klaiber_strongfield_2017,klaiber_relativistic_2023a}, and derived cutoff limits for relativistic rescattering \cite{klaiber_limits_2017}.

In the non-relativistic regime, the effect of spin and spin-orbit coupling was found to play a role in atoms with appreciable internal spin-orbit coupling. Early work demonstrated spin-orbit effects in ion alignment with strong laser fields \cite{santra_spinorbit_2006}.
Spin-orbit effects in particular cases in single- \cite{fano_spin_1969} and multi- \cite{lambropoulos_spinorbit_1973} photon ionization have long been known, but in Ref.~\cite{barth_spinpolarized_2013}, this was theoretically described for strong-field ionization with a circularly polarized laser field (later extended to a two colour field \cite{kaushal_looking_2018a}), where a general mechanism for producing spin polarized electrons was developed.
%that did not require the fine-tuning of the single- and multiphoton regime. 
This mechanism depends on the preferential tunnelling of `counter'-rotating electrons to the circular laser field \cite{barth_nonadiabatic_2011,herath_strongfield_2012, eckart_ultrafast_2018}. The spin polarized electrons were experimentally verified by Ref.~\cite{hartung_electron_2016}, where Mott scattering was used to measure the spin polarization of the photoelectrons. Later experiments revealed different spin-polarization across peaks in the photoelectron ATI spectra \cite{trabert_spin_2018}.
Subsequent theoretical work has focused on effects due to Pauli symmetrization given initial singlet or triplet states \cite{zille_spindependent_2017,milosevic_spindependent_2017}, or the spin dynamics of the residual ion \cite{barth_hole_2014,kaushal_spin_2015,carlstrom_rydberg_2022}, related work using attosecond pulses has demonstrated the possibility to reveal time-resolved spin dynamics in the ion \cite{zhong_attosecond_2020}.
In a very recent work \cite{carlstrom_control_2023}, inelastic recollision of the photoelectron with the ion was also considered, where tuning the evolution of the ionic spin could allow a spin flip, which has implications for recollision based imaging process such as LIED. %\textbf{Should we add a remark about the recent possibility for attoseond physics (a spin-off from strong-field physics to reveal time-resolved spin dynamics in the ion? See, e.g., a work from the Lund group https://www.nature.com/articles/s41467-020-18847-1  I'm not sure since this could make the discussion less focussed} 

Thus, the potential for spin and spin-orbit coupling to play a role in the photoelectron recollision processes has only just begun to be explored. In this work, we will explore the spin and spin-orbit coupling of the recolliding photoelectron in detail. In order to do this, we will extend the accurate and flexible path-integral approach, the Coulomb quantum-orbit strong-field approximation (CQSFA) \cite{lai_influence_2015,lai_nearthreshold_2017,maxwell_coulombcorrected_2017,maxwell_analytic_2018,maxwell_coulombfree_2018,maxwell_treating_2018}. This model has been developed in the spirit of the quantum-orbit formalisms of the SFA \cite{becker_abovethreshold_2002, milosevic_abovethreshold_2006, amini_symphony_2019}, that has enabled significant insight into ATI and recollision processes \cite{paulus_rescattering_1994, lewenstein_rings_1995}. In its original formulation, the CQSFA was used extensively to understand holographic interference patterns that occur in photoelectron momentum distributions (PMDs)  \cite{kang_holographic_2020,maxwell_spirallike_2020,werby_dissecting_2021,werby_probing_2022}. In these studies, a good qualitative agreement with numerical solutions of the time-dependent Schr\"odinger equation (TDSE) was possible for medium photoelectron energies of up to around $3$--$4\up$.
However, recent developments \cite{brennecke_gouy_2020,rodriguez_forward_2023,carlsen_precise_2023} have enabled exceptional quantitative agreement with the TDSE up to the highest rescattering energies of $10\up$. These developments include, the proper computation of the stability prefactor and Maslov phases \cite{brennecke_gouy_2020}, an improved algorithm for finding saddle-point solutions, the inclusion of a $\sin^2$ pulse envelope, and an improved method for computing the bound-state prefactors, all these improvements are combined in Ref.~\cite{carlsen_precise_2023}, and have been used in this work.

The CQSFA provides the ideal platform for understanding the spin-orbit effects of the recolliding photoelectron, providing an intuitive trajectory-based picture. 
In this work, we demonstrate the high level of agreement of the CQSFA for the non-relativisitic spin-less, few-cycle case, via comparison with a TDSE solver at a wavelength of $1600$~nm for hydrogen at a typical intensity of $5 \times 10^{13}$ W/cm$^2$. 
We demonstrate, even at this stage, that relativistic corrections are required to correctly describe the rescattered wavepacket, due to trajectories that reach superluminal velocities. Thus, we present a derivation for the CQSFA that includes the Breit-Pauli relativistic corrections, in particular, a spin-orbit coupling term and corrections to the kinetic energy (or mass correction term). Inclusion of coherent-spin states, allows for a dynamical description of spin with path integrals. 
We derive analytic expressions that describe the spin dynamics and evolution of the spin-orbit phase.
We find that with the kinetic energy corrections, the rescattering dynamics is properly described, with trajectories no longer travelling faster than light. This leads to a noticeable change in the PMD probability that can be explained in terms of the trajectories.
%\textbf{Is this too implicit - I mean do all the readers understand at this point this discussion of 'initial-condition'}.
Furthermore, we consider the effect of initial spin-alignment vs no spin-alignment. If the relativistic kinetic energy corrections are not included, this leads to considerable differences that would be experimentally measurable. However, including the kinetic energy corrections, leads to very modest differences for the two spin alignments. This demonstrates the importance of including the relativistic kinetic energy corrections when computing the effect of spin-orbit coupling. 
Finally, we derive analytical expressions to approximately identify the (back)-scattering angles for which a relativistic treatment is required across a range of laser intensities.

The article is organized as follows. In \secref{sec:Theory} we briefly describe the theory for the non-relativistic spin-less CQSFA and TDSE solver. In \secref{sec:Results}, we present the results of these models and investigate the rescattered trajectories. In \secref{sec:RelativisticCQSFA}, we present the new formulation of the CQSFA, including spin and other relativistic corrections. In \secref{sec:Analytic}, analytical results are presented regarding the weak-coupling approximation and spin-orbit phase. In \secref{sec:RelResults}, we present the results with the newly derived theory, investigating the effect of the relativistic kinetic energy corrections and spin-orbit coupling terms. In \secref{sec:NewLimits}, we derive limits to determine at which scattering angles a relativistic treatment is required. Finally, in \secref{sec:Conclusions}, we present our final conclusions. Atomic units are used throughout unless stated otherwise.

\section{\label{sec:Theory}Theory}
\subsection{CQSFA}
We begin by giving a brief description of the CQSFA, which has been described in detail in its present accurate form in Ref.~\cite{carlsen_precise_2023}, 
%\textbf{Probably it is better to cite only Carlsen here. The rodriguez work was mentioned above together with brennecke .The Carlsen work is really what we use here and hence such citation would be more accurate and helpful for the reader?} 
while its initial development is described in Refs.~\cite{lai_influence_2015,maxwell_coulombcorrected_2017,maxwell_coulombcorrected_2017,maxwell_analytic_2018}.

We start from the non-relativistic Hamiltonian for an atomic target in a strong-field under the single-active-electron approximation, which may be written as
\begin{equation}
    \hat{H}(t)=\hat{H}_0+\hat{H}_{I}(t).
\end{equation}
Here, $\hat{H}_0$ is the Hamiltonian atomic system, which, we write as $\hat{H}_0=K(\pbh)+U(\hat{\rb})$, where $K(\pbh)$ and $U(\hat{\rb})$ are general functions for the kinetic and potential energies, respectively, while $\hat{H}_{I}(t)=\rbh \cdot \Eb(t)$ describes the interaction with the external laser field in the length gauge.
%or $\hat{H}_{I}(t)=\pbh\cdot\Ab(t)+\Ab^2(t)/2$ in the velocity gauge. 
We want to compute the momentum-dependent transition amplitude $M(\pb)=\braket{\psi_{\pb}|U(t,t_0)|\psi_0}$, where $\ket{\psi_0}$ is the initial bound state of the system, $\ket{\psi_{\pb}}$ is a scattering state with asymptotic momentum $\pb$, the final time $t\rightarrow\infty$, and the initial time $t_0\rightarrow-\infty$. The transition amplitude may be written, still in an exact form, \cite{lai_influence_2015,maxwell_coulombcorrected_2017} as
\begin{equation}
    M(\pb)=-i\int_{-\infty}^{t} d t' \braket{\psi_{\pb}|U(t,t')H_I(t')|\psi_0(t')},
\end{equation}
while insertion of the resolution of the identity operator before $H_{I}(t')$, given by $\iden_{3}=\int d^3 \pb_0 \ket{\pbt_0}\bra{\pbt_0}$, where $\tilde{\pb}_0=\pb_0+\Ab(t')$, enables representation in path-integral form
\begin{align}
    M(\pb)&=-i\int\displaylimits_{-\infty}^{\infty} d t' 
    %\hspace{-2mm}
    \int\displaylimits_{\rb(t')} \frac{\mathcal{D}'\rb}{(2\pi)^3}
    %\hspace{-2mm}
    \int\displaylimits^{\pb(t)}  \mathcal{D}'\pb 
    e^{i S[\rb,\pb,t']}d(\tilde{\pb}_0,t'),
    \label{eq:Mp}
\intertext{with $d(\tilde{\pb}_0,t')=\braket{\pb_0+\Ab(t')|H_I(t')|\psi_0|}$,}
S[\rb,\pb,t']&=\notag\\
\ip t'&-\pb(t')\cdot\rb(t') 
-\frac{1}{2}\int_{t'}^{\infty} d\tau 
\left(
    \dot{\pb}\cdot\rb+H[\rb,\pb,t']
\right)\label{eq:Action},
\end{align}
and $H[\rb,\pb,t']$ is the classical Hamiltonian \footnote{Technically $H[\rb,\pb,t']$ is the Weyl transformed quantum-mechanical Hamiltonian \cite{mizrahi_weyl_1975}, which differs from the classical Hamiltonian $H[\pb,\rb,t']=K[\pb]+U[\rb]+H_I(t')$ by orders of $\hbar^2$, however these differences may be discarded given we will be applying the saddle-point approximation that already neglects the quadratic power of $\hbar$.}. The action in \eqref{eq:Action} results from enforcing an initial condition in position-space and a final limit in momentum space, giving rise to the boundary term $\pb(t')\cdot\rb(t')$, i.e., we are in the mixed representation. These restrictions are denoted by the primes on $\mathcal{D}$'s, full details are given in Ref.~\cite{carlsen_precise_2023}. 
For atoms, the initial position is the origin, hence $\rb(t')=\mathbf{0}$, and $\pb(t')\cdot\rb(t')$ may be dropped. Note, we define $\pb_0=\pb(t')=\pb(\Re[t'])$, which results from taking momentum fixed during tunnelling, this approximation is explained in more detail in Appendix \ref{Sec:AppendixCQSFA}.

The transition amplitude of \eqref{eq:Mp} is then evaluated via the saddle-point approximation
\begin{equation}
    M(\pb)=-i\sum_s \sqrt{\frac{2\pi i}{\partial^2 S/\partial t'^2}}
    \frac{e^{-i\pi\nu/2}}{\sqrt{|J|}}
    d(\tilde{\pb}_{0s},t_s)e^{i S[\rb_s,\pb_s,t_s]},
    \label{eq:TransSaddle}
\end{equation}
where $\underline{\underline{J}}=\frac{\partial \pb_f}{\partial \pb_0}$, $J=\det(\underline{\underline{J}})$, and $\nu$ is the Maslov phase, which may be determined by computing $\underline{\underline{J}}$ at all points in time and counting the number of focal points ($J=0$) 
%\textbf{I tried to introduce the matrix notation with two underlines. Please chek the meaning of the condition that you wirte J=0, do you mean that the determinant should be zero?} 
\cite{carlsen_precise_2023,brennecke_gouy_2020}. 
The sum runs over solutions to the saddle-point equations, which are given by
\begin{align}
(\pb_{0s}+\Ab(t_s))^2+2\ip=0\label{eq:IonizationTimes}\\
\dot{\rb}_s(t)=\nabla_{\pb}K[\pb_s] \quad
\dot{\pb}_s(t)=-\nabla_{\rb}U[\rb_s]. \label{eq:NewtEqns}
\end{align}
Equation (\ref{eq:IonizationTimes}) leads to complex ionization times $t_s$ meaning the integral in \eqref{eq:Action} is done in two parts, first over imaginary time from $t_s$ to $\Re[t_s]$ associated with tunnelling, then over real time from $\Re[t_s]$ to $\infty$, associated with real-space continuum propagation, see Appendix \ref{Sec:AppendixCQSFA} for more details. 

The CQSFA solves a boundary value problem known as the inverse problem, i.e., 
all solutions of Eqs.~(\ref{eq:IonizationTimes}) and (\ref{eq:NewtEqns}) (a.k.a. trajectories or quantum orbits) are found satisfying $\rb(t')=\mathbf{0}$ and $\pb(t\rightarrow\infty)=\pb_f$. There will be multiple solutions for each final momentum point, each solution can be uniquely determined by its initial momentum coordinate $\pb_0$. Thus, the inverse problem reduces to finding the set of initial momentum coordinates $\pb_0$, for each final momentum $\pb_f$.

The approach just described is different from the majority of models, which use a forward approach and bin trajectories with similar final momenta, which requires small bins and many trajectories to resolve interferences, while the inverse approach allows for many fewer trajectories. Furthermore, recent work \cite{shvetsov-shilovski_semiclassical_2021} has identified that forward approaches do not yield the correct sampling weight in terms of the Jacobian $J$, leading to $1/|J|$ instead of the correct $1/\sqrt{|J|}$ computed by inverse approaches.
The downside of the inverse approach is that it is much harder to solve, and approaches can be less general.
In previous works \cite{maxwell_coulombcorrected_2017,maxwell_coulombfree_2018,maxwell_analytic_2018}, solutions were found by `exploring' the manifold of solutions, i.e.,  changing parameters to find `connected' solutions. This method is very fast but has two disadvantages; (i) it requires preknowledge of the shape of the manifold and, where solutions lie, which changes from system to system; (ii) it assumes all solutions can be reached in this way, which is not the case. In this work, we used an alternative method \cite{carlsen_precise_2023} where the solutions are found by initial random sampling of $\pb_0$, and adaptively concentrating guesses in $\pb_0$ regions where dense clusters of solutions are found. This method is more general and allows all solutions to be found regardless of the system.

\begin{table}
\centering
\begin{tabular}{ c c c l}
  \hline
  Orbit & $\Pi_{z}$ & $\Pi_{\perp}$ & Behaviour\\%$z_0 p_{fz}$ & $p_{fx}p_{0z}$ & Behavior\\ 
  \hline\hline
  1 & + & + & Direct \\ 
  2 & - & + & Laser-driven deflection \\ 
  3 & - & - & Forward Scattered \\ 
  4 & + & - & Rescattered \\ 
  \hline
\end{tabular}
\caption{Orbit classification used in the CQSFA for monochromatic linearly polarized fields. The labelling 1 to 4 classifies the orbit with two different conditions, the signs of $\Pi_{z}=z_0p_{fz}$ and $\Pi_{\perp}=p_{fx}p_{0x}+p_{fy}p_{0y}$, respectively. The behaviour in the fourth column indicates the expected dynamics of the specific types of orbits.  }
\label{tab:OrbitClassification}
\end{table}

The distinct solutions for each $\pb_f$ do not have an exhaustive classification, however, in the case of linearly polarized fields, four broad types of orbit can be defined that can helpfully classify behaviour. This classification uses the tunnel exit 
\begin{equation}    z_0=\Re\left[\int_{t_s}^{\Re[t_s]}\!\!d\tau A_z(\tau)\right],
\end{equation}
and initial momentum perpendicular to the laser polarization $\pb_{0 \perp}$, and compares their sign with $p_{fz}$ and $\pb_{f\perp}$, respectively. Then we may define $\Pi_{z}=z_0p_{fz}$ and $\Pi_{\perp}=p_{fx}p_{0x}+p_{fy}p_{0y}$. This is the same definition as in Ref.~\cite{carlsen_precise_2023}, which generalizes the commonly employed 2D classification \cite{yan_lowenergy_2010, maxwell_coulombcorrected_2017,rodriguez_forward_2023} to 3D. The classification, summarized in Table \ref{tab:OrbitClassification}, is as follows; $\Pi_{z}>0$ $\land$ $\Pi_{\perp}>0$: orbit 1, associated with direct trajectories that do not return to the parent atom;
$\Pi_{z}<0$ $\land$ $\Pi_{\perp}>0$: orbit 2, describes trajectories that undergo a laser driven return but do not interact strongly with the parent atom;
$\Pi_{z}<0$ $\land$ $\Pi_{\perp}<0$: orbit 3,  trajectories that forward scatter off the parent atom; and 
$\Pi_{z}<0$ $\land$ $\Pi_{\perp}<0$: orbit 4, associated with trajectories that back-scatter off parent atom. Note, that there may be more than one valid solution of each type. 

\begin{figure*}
    \centering  
    \includegraphics[width=0.85\textwidth]{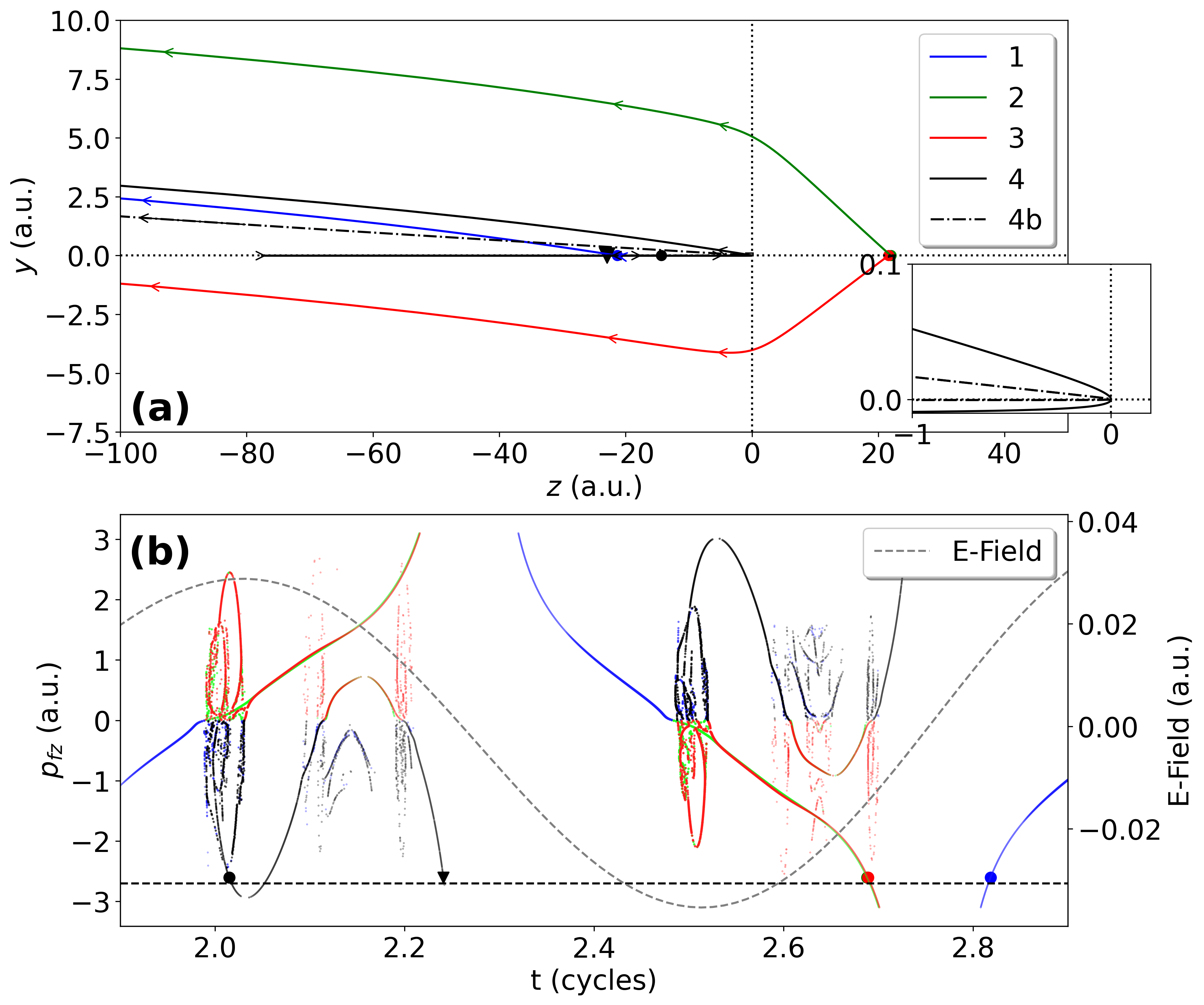}
    \caption{Example trajectories for all 4 orbits (a), and the times of ionization for different final momenta $p_{fz}$ (b).  
    In the case of orbit 4 two solutions are shown, one “regular” laser driven recollision, and one “directly” recolliding, labelled 4b. The laser is linearly polarized along the $z$ axis. The tunnel exits are marked by circles (triangle for 4b).
    (a) shows the four orbit types, the $y$-axis has been stretched to better
    show the detail. (b) plots the real time of ionization vs the $p_z$ momentum over a single laser cycle, with the 6-cycle electric field superimposed. The ionization target is hydrogen, the laser wavelength is $\lambda=1600$~nm and the intensity is $I_0=5\times10^{13}$~W/cm$^2$.
    %The point size and opacity is scaled with the probability, so that higher probability points are more visible.
    An inset showing the dynamics of orbit 4 and 4b close to the core is given, with its region marked by the rectangle (a).
    The final perpendicular momentum is chosen to be $p_{fx}=0.0$~a.u. $p_{fy}=0.05$~a.u., while in panel (a) $p_{fz}=-2.704$, which is denoted by the black dashed line in panel (b). 
    Note the ionization times for orbit 2 in (b) are almost the same as those for orbit 3, and thus mostly obscured.
    The solutions corresponding to the trajecotries plotted in (a) are marked on (b) by circles (triangle for 4b) on the dashed line.
    %The field parameters and final momenta correspond to the horizontal line given on \figref{fig:TDSE_CQSFA}(b). 
    }
    \label{fig:TrajectoriesTimes}
\end{figure*}

\begin{figure*}
    \centering  
    \includegraphics[width=\textwidth]{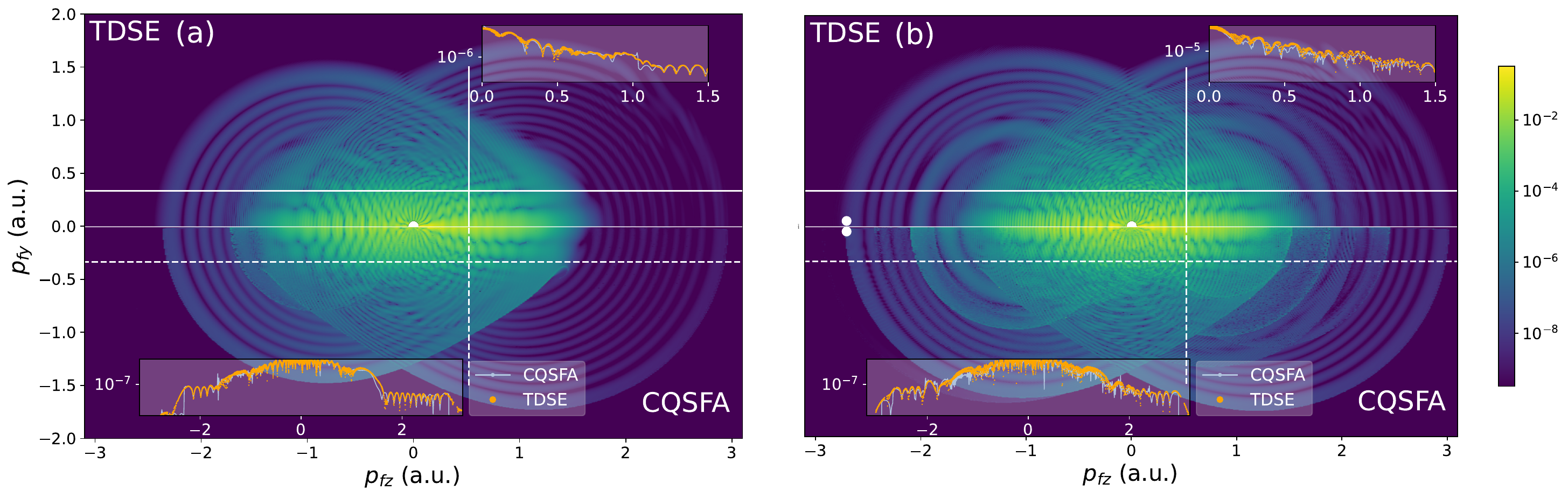}
    \caption{Comparison of PMDs for strong-field ionization of hydrogen computed using the TDSE and CQSFA, for a 4-cycle and 6-cycle sin$^2$ pulse, panels (a) and (b) respectively. The upper half of each panel is computed using the TDSE solver \Qprop{}, while the lower half is computed using the CQSFA. The insets show lineout plots of the TDSE and CQSFA PMDs. With the white horizontal lines showing the region plotted in the bottom insets, and the white vertical lines showing the region plotted in the top insets. The intensity is $I_0=5\times10^{13}$~W/cm$^2$ and wavelength $\lambda=1600$~nm, the target is hydrogen with $\ip=0.5$~a.u. The white dots to the left in panel (b) correspond to the example trajectories plotted in Figs.~\ref{fig:TrajectoriesTimes} and \ref{fig:RelSurprise}.}
    \label{fig:TDSE_CQSFA}
\end{figure*}

Examples typifying the classifications are given in \figref{fig:TrajectoriesTimes}(a); laser parameters are given in the caption. Here we see typical examples of orbits 1-4, as well as an `atypical' orbit 4 (labelled 4b), along with the corresponding ionization times in the panel below.
For this study, we are most interested in the rescattered trajectories of orbit 4.  The tunnel exit of the `typical' orbit 4 lies at $z_0\approx-15$~a.u., from here the laser drives the trajectory away and then back again, undergoing a laser driven return, as shown in \figref{fig:TrajectoriesTimes}(a). Thus, this solution belongs to a well-studied category of returning trajectory known from HHG and ATI \cite{lewenstein_theory_1994,paulus_rescattering_1994}. These orbits come in pairs, long and short. Here, we have only plotted the long trajectory as the short's path would be very similar. These pairs can be seen in \figref{fig:TrajectoriesTimes}(b) in every half cycle by black loops extending in opposite directions each half cycle. The solution pairs form loops because at a classical boundary, they coalesce, forming the rescattering ridge in the PMD, see \figref{fig:TDSE_CQSFA}. In contrast, the directly recolliding orbit 4b has a tunnel exit of $z_0\approx-20$~a.u.\ close to orbit 1's tunnel exit. Instead of undergoing a laser driven recollision the initial conditions are such that the trajectory returns before the laser has considerably changed sign. These trajectories also occur every half cycle and are connected to the `typical' recolliding orbits via orbit 3 on a continuous manifold. Orbit 4b ionizes far from the peak at near zero field, thus the probability of these orbits is very low, however, they undergo a very strong recollision, so provide an interesting case to study. If the probability of these trajectories could be increased, they could be used for very strongly probing targets.

\subsection{TDSE}
To benchmark the CQSFA results, we solve the TDSE using the freely available \Qprop{} \cite{tulsky_qprop_2019} software.
\Qprop{} is a single-active-electron TDSE solver, which implements a fast and accurate method for the calculation of PMD using the i-SURFF projection method. To model hydrogen, a Coulomb potential is employed. In this computation, we considered angular momenta up to $l=200$, grid spacing $\Delta r = 0.1$~a.u., and time step $\Delta t = 0.05$~a.u, and we checked the results for convergence.

In both the CQSFA and \Qprop{} we consider a sin$^2$ laser field, where the vector potential is defined by
\begin{equation}
        \Ab(t)=2\sqrt{\up}\sin^2\left( \frac{\omega t}{2 N}\right)\cos(\omega t + \phi),
\end{equation}
with $N$ being the number of laser cycles of the vector potential envelope, while $\up$ is the ponderomotive energy or quiver energy of the free electron in the laser field, which is proportional to the peak laser intensity $I_0=2\up c \epsilon_0 \omega^2$. The angular frequency is given by $\omega$ and the carrier envelope phase (CEP) is given by $\phi$. We focus on a wavelength $\lambda=1600$~nm, i.e., $\omega=0.0285$~a.u.

\section{\label{sec:Results}Non-Relativistic Results}
%\subsection{\label{sec:Non-Relativisitc}Non-Relativistic}
In \figref{fig:TDSE_CQSFA}, we show a validation of the CQSFA vs the non-relativistic-TDSE with PMDs for hydrogen at a wavelength of $1600$~nm for a 4-cycle and 6-cycle sin$^2$ pulse, panels (a) and (b), respectively. Very good agreement is seen between the two models across the whole momentum region. There is some deviation across classical boundaries at higher energies, seen in the lineout insets, where there is a more abrupt change from higher signal to lower signal as the classical boundary is crossed. Improved agreement here can only be achieved by accounting for correctly treating the coalescing long and short orbit 4 solutions, seen in \figref{fig:TrajectoriesTimes}, in a manner similar to the uniform approximation \cite{figueirademorissonfaria_highorder_2002} applied to the equivalent SFA orbit, see e.g., \cite{kocia_semiclassical_2016}.

The high-energy rings visible in \figref{fig:TDSE_CQSFA}(a) [around $(p_{fz},p_{fy})=(-2.0, 0.0)$] and \figref{fig:TDSE_CQSFA}(b) [around $(p_{fz},p_{fy})=(-2.5, 0.0)$] are particularly well-captured. Previously these have not be very well described by the CQSFA, but now additional orbit 4 contributions have been included that lead to very good agreement.
The high-energy rings can be completely attributed to `typical' pairs of orbit 4, where the interference between pairs gives the ring-like interference, previously only the long orbit was included in the CQSFA. 
%This interference structure in different regions can be associated with specfic orbit 4 pairs.
In \figref{fig:TDSE_CQSFA}(b), the left most high-energy rings [around $(p_{fz},p_{fy})=(-2.5, 0.0)$] can be attributed to the orbit 4 pairs ionized near the peak at 2 cycles (\figref{fig:TrajectoriesTimes}(b)), while the right most high-energy rings [around $(p_{fz},p_{fy})=(3.0, 0.0)$] can be attributed to the orbit 4 pairs ionized near the peak at 2.5 cycles, see \figref{fig:TrajectoriesTimes}(b).

The general high level of agreement seen in \figref{fig:TDSE_CQSFA} validates the CQSFA approach, allowing the use of the trajectory-based machinery for interpretation and to probe the physics in detail. This also provides an easy platform to extend the formalism to include additional effects, such as the spin-orbit coupling during photoelectron recollisions.

In \figref{fig:RelSurprise}, we plot metrics for five example high-energy rescattered trajectories following ionization of hydrogen for the 6-cycle laser pulse, in preparation for including spin-orbit coupling dynamics, this includes the evolution of the velocity [\figref{fig:RelSurprise}(a)], distance from the origin [\figref{fig:RelSurprise}(b)], and the kinetic energy, potential and spin-orbit phase terms in the action [\figref{fig:RelSurprise}(c)]. We choose two pairs of rescattered trajectories that are ionized near the field peaks at 2 and 3 laser cycles [see ionization times marked on the laser field in \figref{fig:RelSurprise}(a)] and recollide around field crossings after 2/3 to 3/4 of a cycle of propagation, note this differs from the well-known 2/3 result \cite{paulus_rescattering_1994} due to the inclusion of the Coulomb field. These trajectories are the previously discussed orbit 4 pairs,  responsible for the high-energy ring-like structures in \figref{fig:TDSE_CQSFA}. We also include orbit 4b in \figref{fig:RelSurprise}. As previously mentioned orbit 4b has a low probability, so no clear features are visible in the spectrum in \figref{fig:TDSE_CQSFA}, however, it undergoes a very high-energy recollison so is a useful test case.

Despite employing non-relativistic intensities and all final photoelectron speeds being far from relativistic, the momentum transfer during recollision is strongly relativistic. This can be seen in \figref{fig:RelSurprise}(a), by the large superluminal velocity spikes, occurring during recollisions. The long and short orbit 4 trajectories consistently reach $v\approx 2 c$, while for the directly recolliding orbit $v\approx 6 c$. These fast speeds are due to how strongly the core is probed, \figref{fig:RelSurprise}(b) shows that the trajectory's distance from the origin goes below $10^{-5}$~a.u. 
Therefore, relativistic considerations have an impact for a sizeable part of the recolliding wavepacket, that, as we will show, noticeably affects the PMD in regions dominated by rescattering.
However, these high speeds are only reached for very short time-periods, within a single attosecond. Thus, if it is for such short times, can it be simply neglected?

The phases in \figref{fig:RelSurprise}(c), where we plot the time-integral of the kinetic energy $K(t)$, potential energy $V(t)$, and spin-orbit interaction $H_{SO}(t)$, see \eqref{eq:Rel_length}, help us to understand the importance of these very fast speeds over very short time-periods. Given that the atomic core is probed so strongly, a large phase due to spin-orbit coupling is acquired. In the case of the directly recolliding trajectory, the spin-orbit phase (in purple) actually exceeds the phase picked up from the Coulomb potential. This is unexpected and requires careful analysis. 
It seems likely that these results may be overestimated, while they include trajectories with superluminal speeds. Crucially, the spin-orbit phase only really appreciably changes in the attosecond time-period when superluminal velocities are reached. This can be seen from the highly abrupt change in the dashed-dotted lines in \figref{fig:RelSurprise}(c).

Overall, in \figref{fig:TDSE_CQSFA}, there is very good agreement with the TDSE, but this is the non-relativistic solutions of the \Schr equation, which allows for portions of the wavepacket to travel faster than the speed of light, as we see here. 
Thus, we can deduce that there is a non-trivial contribution from rescattered wavepackets, which have a superluminal group velocity.
Clearly this is unphysical, and therefore we should go beyond the TDSE and nonrelativistic CQSFA.
%For a correct treatment, we should look to the Dirac equation and relativistic corrections terms to the standard TDSE. 

\section{\label{sec:RelativisticCQSFA}Formulation of Relativistic CQSFA}
\begin{figure*}
    \centering
    \includegraphics[width=\linewidth]{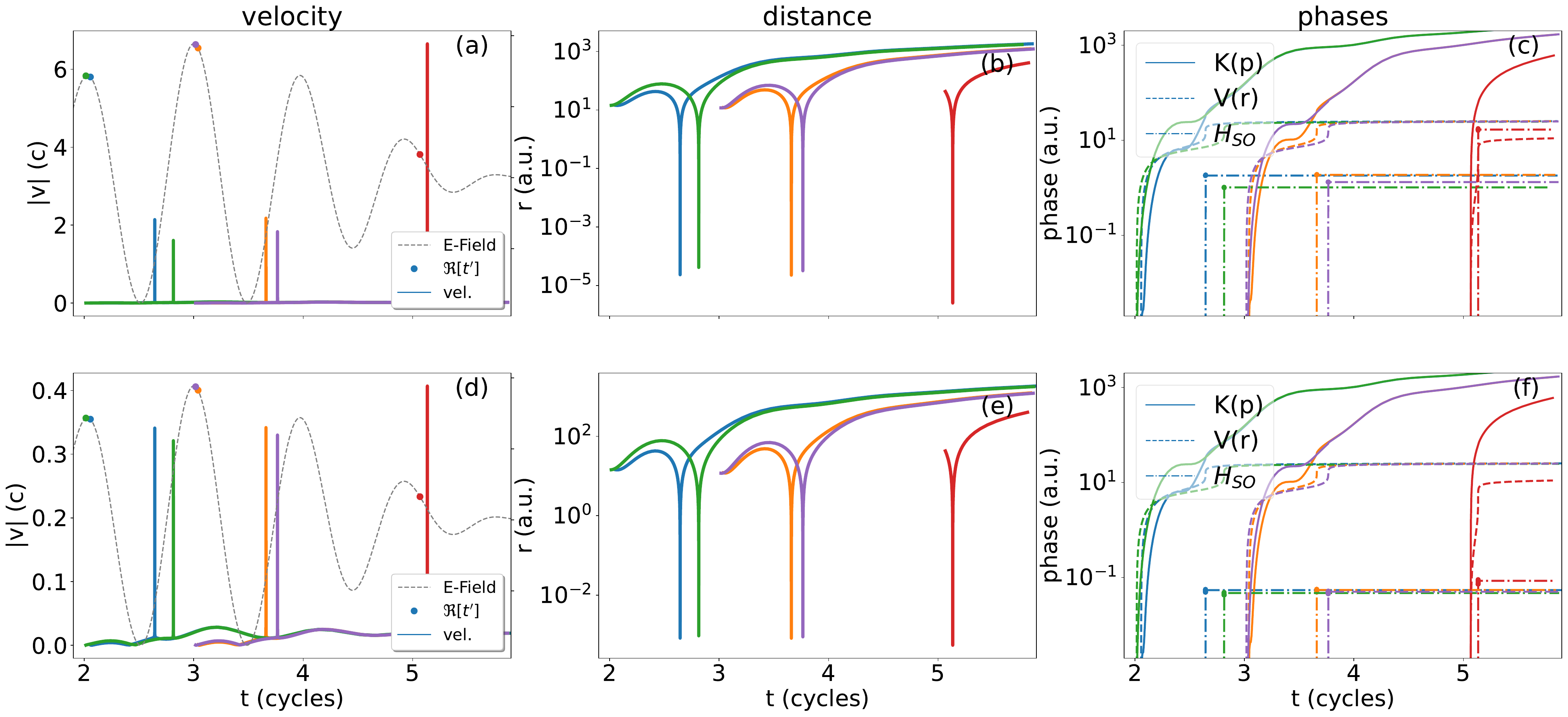}
    \caption{Examples of velocity transfer and resulting phases picked up of rescattered orbit 4 trajectories, following strong-field ionization of hydrogen. 
    (a) and (d):  Velocity over time given by coloured lines (with abrupt spikes at recollisions) with the laser field plotted by a black dashed line, where ionizations times are marked on the field by coloured circles, where the colour of the circle and velocity line corresponds to particular orbits. Blue and green correspond to long and short trajectories with an ionization time around 2 cycles, orange and purple, to long and short trajectories with an ionization time around 3 cycles, and red, to an orbit 4b trajectory with an ionization time around 5 cycles.
    (b) and (e): Distance over time given by coloured lines, the colours match the same orbits as panel (a). 
    (c) and (f): The phase acquired from kinetic energy $K(p)$ [solid lines], potential energy $U(r)$ [dashed lines], and spin-orbit coupling $H_{SO}$, see Eqs.~(\ref{eq:Hso}) and (\ref{eq:Cso}), [dot-dashed lines with an abrupt increase during recollision], which is given by the time-integral of these quantities over each trajectory. 
    The approximation to the spin-orbit phase, given by \eqref{eq:SpinOrbit-Analytic}, is plotted in (c) and (f) as coloured circles, as before the colours match the orbits in panel (a).
    The final momentum of all trajectories is the same as \figref{fig:TrajectoriesTimes} $\pb_f=(0.0, 0.05, -2.704)$, this point is indicated on \figref{fig:TDSE_CQSFA}. }
    \label{fig:RelSurprise}
\end{figure*}

\begin{figure}
    \centering
    \includegraphics[width=\linewidth]{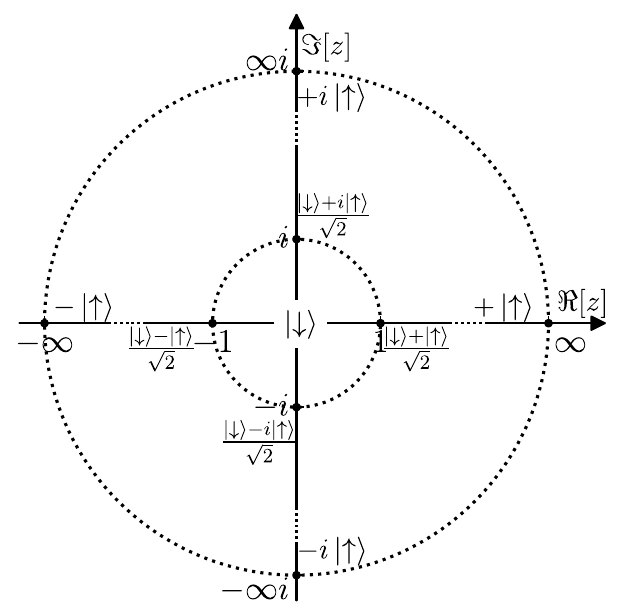}
    \caption{The mapping of the complex plane to coherent spin states $\ket{z}$, given by \eqref{eq:spin-state}. Some key values are marked by black dots and their corresponding states are given by adjacent labels. }
    \label{fig:CoherentSpin}
\end{figure}

%\subsection{\label{sec:RelativisticTheory}Relativistic Theory}
The results in \figref{fig:RelSurprise}, show that the most rescattered part of the wavepacket, gains unexpected large velocities, which warrants a relativistic treatment including spin-orbit coupling. The starting point for this is the Dirac equation, but as we are not considering relativistic laser intensities we will consider the expansion of the Dirac equations, up to terms of order $(v/c)^2$ \cite{strange_relativistic_1998}. 
%Then we may also write this as correction terms to the non-relativistic case. 
The laser field may be included through minimal coupling, this gives what is sometimes referred to as the Breit-Pauli Hamiltonian
\begin{align}
    \hat{H}(t)&=\frac{1}{2}\left(\pb+\Ab(\hat{\eta})\right)^2
    +\hat{\Sb}\cdot\Bb(\hat{\eta}) 
    -\frac{1}{8 c^2}\left(\pbh+\Ab(\hat{\eta})\right)^4\notag\\
    &+V(\rbh)-\frac{1}{8c^2}\nabla_{\rb}^2V(\rbh)
    +\frac{\frac{\partial}{\partial r}V(|\rbh|)}
    {2 c^2 |\rbh|}\hat{\Lb}\cdot\hat{\Sb},
\end{align}
where $\hat{\eta}=\omega t - \kb\cdot\rbh$, $\Ab(\hat{\eta})$ and $\Bb(\hat{\eta})$, are the magnetic vector potential and magnetic field, respectively, while $\hat{\Lb}=\rbh\times(\pbh+\Ab(\hat{\eta}))$ is the angular momenta, $\hat{\Sb}=\hat{\sigma}/2$ is the spin operator, and $\hat{\sigma}=\sigma_x \hat{x} + \sigma_y \hat{y} + \sigma_z \hat{z}$ is the Pauli vector, constructed from the Pauli matrices.

Given that the wavelength of the laser-field is not too long (or too short) \cite{reiss_limits_2008} we may apply the dipole approximation, which allows the velocity and length gauge forms of the Hamiltonian % $\hat{\tilde{\mathbf{p}}}$
\begin{align}
    \hat{H}_v(t)&=K(\pbth)+U(\rbh)+\mathbf{C}_{SO}(\rbh,\pbth)\cdot\hat{\Sb}\\
    \hat{H}_l(t)&=
    \underbrace{K(\pbh)+U(\rbh)}_{\hat{H}_0}
    +\underbrace{\mathbf{C}_{SO}(\rbh,\pbh)\cdot\hat{\Sb}}_{\hat{H}_{SO}}
    +\hat{H}_I(t)\label{eq:Rel_length}
    \intertext{with}
    K(\pbh)&=K_0(\pbh)+K_1(\pbh)=\frac{\pbh^2}{
    2}-\frac{\pbh^4}{8c^2}\label{eq:K-Corr}\\
    U(\rbh)&=U_0(\rbh)+U_1(\rbh)=V(\rbh)-\frac{\nabla^2_{\rb}V(\rbh)}{8 c^2}\\
    \mathbf{C}_{SO}(\rbh,\pbh)&=\frac{\frac{\partial }{\partial r}V(|\rbh|)}{2c^2|\rbh|}\rbh\times\pbh
\end{align}
and $\hat{H}_{I}(t)=\Eb(t)\cdot\rbh$ and $\pbth=\pbh+\Ab(t)$.
The relativistic correction terms to the kinetic and potential energy (the Dyson correction), $K_{1}(\pbh)$ and $U_1(\rbh)$, still permit the solution given in \eqref{eq:TransSaddle}, as here we assumed general forms for the kinetic and potential energy. The spin and spin-orbit coupling can, however, not be treated through the same path-integral approach that we used before, as we must consider the additional spin degree of freedom and coupling between them. In fact, historically a proper treatment of spin by path-integrals took some time to develop.
In order to describe spin in a path-integral framework, we require a mapping of the spin to a continuous variable. The treatment of spin in terms of so-called coherent spin-states (an irreducible representation for $SU(2)$) \cite{klauder_path_1979,pletyukhov_semiclassical_2002} does just this.
%spin-orbit coupling term, however, calls for a treatment of spin, and it's coupling to angular momenta. %Thus, in the next section, we will describe how to build spin into the path-integral formalism.
The mapping is achieved through the coherent spin states, defined as
\begin{equation}
 \ket{z}=\frac{\ket{\downarrow}+z\ket{\uparrow}}{\sqrt{1+|z|^2}},
 \label{eq:spin-state}
\end{equation}
with further details given in Appendix \ref{sec:SpinStateFrame}. These states map spin states to the complex plane, 
 see \figref{fig:CoherentSpin}. Some key values are $\ket{z\rightarrow0}=\ket{\downarrow}$, $\ket{z\rightarrow\infty}=\ket{\uparrow}$, $\ket{z\rightarrow1}=\ket{+}$ and $\ket{z\rightarrow-1}=\ket{-}$. We may write an initial-fine-structure state \cite{bransden_physics_1983}, with quantum numbers $j$, and $m_j$, and energy $E_{j m_j}$ as
\begin{equation}
	\ket{\Phi_{j m_j}}=\sum_{m} f^{j m_j}_{l m} \ket{z^{j m_j}_{l m}}\otimes\ket{\psi_{l m}}
        \label{eq:spin-state1}
\end{equation}
with $f^{j m_j}_{l m}=C^{j m_j}_{l m, \frac{1}{2} -\frac{1}{2}}
\sqrt{1+|z^{j m_j}_{l m}|^2}$, $z^{j m_j}_{l m}=C^{j m_j}_{l m, \frac{1}{2} \frac{1}{2}}/C^{j m_j}_{l m, \frac{1}{2} -\frac{1}{2}}$ and $C^{j m_j}_{l m, S m_s}=\braket{l,S;m,m_s|l,S;j,m_j}$ are Clebsch-Gordon coefficients.

%\subsection{\label{sec:SpinTransAmp}Transition amplitude with spin}
The transition amplitude can be defined for ionization starting in a specific bound state $\ket{\Phi_{j m_j}}$ with quantum numbers $j$ and $m_j$ and energy $E_{j m_j}$, and finishing in a continuum spin-state $\ket{\psi_{\pb},z}=\ket{\psi_{\pb}}\otimes\ket{z}$, at $t\rightarrow\infty$. The amplitude is then given by
\begin{align}	
M(\pb,z)&=
\braket{\psi_{\pb},z|U(t,t_0)|\Phi_{j m_j}}\notag\\
&=-i\!\!
\int_{-\infty}^{t}\!\!\!\! d t' \!\!\braket{\psi_{\pb},z|U(t,t')H_I(t')|\Phi_{j m_j}}e^{-i E_{j m_j} t'}.
\end{align}
%\textbf{fix some of these equations on this page such that they fit into one column}

%\textbf{In this column, I suggest that all the tildes ontop of p0 is centered at the 'p' and not at 'p0'}
As before, we insert the resolution of the identity $\iden_{3}=\int d^3 \pbt_0 \ket{\pbt_0}\bra{\pbt_0}$, where $\pbt_0=\pb_0+\Ab(t')$
\begin{align}
M(\pb,z)=-i
%\lim_{\substack{t\rightarrow\infty\\t_0\rightarrow-\infty}}
\int_{t_0}^{t}\! d t' \!\!\int d^3 \pbt_0
&\braket{\psi_{\pb},z|U(t,t')|\pbt_0}
\notag\\
&\times\braket{\pbt_0|\hat{H}_I(t')|\Phi_{j m_j}} e^{-i E_{j m_j} t'}.
\end{align}
To proceed, we must expand the spin state, so we can exploit the fact that $[\hat{H}_{\mathrm{I}}(t),\hat{\Sb}]=0$.
\begin{align}
	&\braket{\pbt_0|\hat{H}_{\mathrm{I}}(t')|\Phi_{j m_j}}
 =\sum_{m}\ket{z^{j m_j}_{l m}}d_{m}(\pbt_0,t')
 %    &=e^{i E_{j m_j} t'} \sum_{m}
	% \overbrace{f^{j m_j}_{l m} \braket{\pb_0|\hat{H_{\mathrm{I}}}|\psi_{l,m}}}^
	% {d_m(\pb_0,t')}
	% \otimes\ket{z^{j m_j}_{l m}}\\
	% &
	\end{align}
with $d_m(\pb_0,t')=f^{j m_j}_{l m} \braket{\pb_0|\hat{H_{\mathrm{I}}}|\psi_{l,m}}$. 
Now the transition amplitude may be written
\begin{align}
M(\pb,z)&=-i\sum_{m}\int_{-\infty}^{t} d t' \mathcal{K}_{m}(\pb,z)e^{-i E_{j m_j} t'},\\
\intertext{where}
\mathcal{K}_m(\pb,z)&= \int d^3 \pbt_0
\braket{\psi_{\pb}(t),z|U(t,t')|\pbt_0,z^{j m_j}_{l m}}d_{m}(\pbt_0,t').
\label{eq:Kernel}
 \end{align}

Now we are in a position to utilize the path-integral formalism for a particle with spin \cite{kochetov_su_1995,pletyukhov_semiclassical_2002,pletyukhov_semiclassical_2003,morten_path_2008} and obtain
\begin{align}
 \mathcal{K}_m(\pb,z)&=\notag\\ 
        \int\displaylimits_{\rb_0} \frac{\mathcal{D}'\rb}{(2\pi)^3} &
        \int\displaylimits^{\pb_f}  \mathcal{D}'\pb
        \int\displaylimits_{z^{j m_j}_{l m}}^{z} \mathcal{D}\mu(z)
        d_{m}(\pb_0,t')
	e^{i \mathcal{A}[\rb,\pb,z,t']}.
 \label{eq:K-PathIntegral1}
 \end{align}
 Here it is possible to solve the functional integral over $z$ analytically. We take advantage of this fact and rewrite \eqref{eq:K-PathIntegral1} as
\begin{align}
    &\mathcal{K}_m(\pb,z)=\notag\\ 
        &\int\displaylimits_{\rb_0} \frac{\mathcal{D}'\rb}{(2\pi)^3} 
        \int\displaylimits^{\pb_f}  \mathcal{D}'\pb
        %\int\displaylimits_{z^{j m_j}_{l m}}^{z} \mathcal{D}\mu(z)
        \mathcal{M}^{m}_{\mathrm{SO}}(\rb,\pb,z,t')
        d_{m}(\tilde{\pb}_0,t')
	e^{i S[\rb,\pb,t']}
 \label{eq:K-split}
\intertext{with}
    &\mathcal{M}^{m}_{\mathrm{SO}}(\rb,\pb,z,t')=\notag\\
    &\frac{a^*(t)-b^*(t)z^{j m_j *}_{l m}+b(t)z+a(t)z^* z^{j m_j}_{l m}}
    {\sqrt{1+|z|^2}\sqrt{1+|z^{j m_j}_{l m}|^2}}.
    \label{eq:SpinOrbitTransition}
\intertext{and}
&S[\rb,\pb,t']=-E_{j m_j} t' -\int_{t'}^{t}d\tau(\dot{\pb}\cdot\rb +H_0[\rb,\pb+\Ab(\tau)]).
\end{align}
In \eqref{eq:SpinOrbitTransition} $a(t)$ and $b(t)$, are time-dependent functions determined by an ordinary differential equation given the appendix \ref{sec:weak-coupling} in \eqref{eq:ab-ode}. Thus, \eqref{eq:K-split} may be solved using the saddle-point equations given by Eqs.~(\ref{eq:IonizationTimes}) and (\ref{eq:NewtEqns}), leading to the final expression for the transition amplitude
\begin{align}
    &M(\pb,z)=\notag\\
    &=-i\sum_{m,s}
    C_m(\rb_s,\pb_s,t_s)
    \mathcal{M}^{m}_{\mathrm{SO}}(\rb_s,\pb_s,z,t_s)
    e^{i S[\rb_s,\pb_s,t_s]}
    \label{eq:RelTransSaddle}
    \intertext{with}
    &C_m(\rb_s,\pb_s,t_s)=\sqrt{\frac{2\pi i}{\partial^2 S/\partial t'^2}}
    \frac{e^{-i\pi\nu/2}}{\sqrt{|J|}}
    d_m(\tilde{\pb}_{0s},t_s).
    \label{eq:RelPref}
\end{align}
Note that this formulation assumes that $\mathcal{M}^{m}_{\mathrm{SO}}$ varies slowly enough so that it does not affect the saddle-point equation for $\rb_s$ and $\pb_s$, which is the weak-coupling limit. Full details of the weak-coupling limit and how we may go beyond it are given in Appendix \ref{sec:weak-coupling} and \ref{sec:Mod-SPA}, respectively.

\section{Observables and analytical considerations}
\label{sec:Analytic}
\subsection{Spin Measurement}

%\textbf{Work with the equations in this section to make them fit into a single column}

%\textbf{In this section, we should take care that the notation is consistent with that used in captions and in the appendix. IN some places a $\mathcal{P}$ is used, in other places just a regular $P$. Probably you have some thoughts about this and I would like you to check these notations carefully in main text and appendix. (I am challenged by the fact that I don't have a printout...)}

So far we have only considered the case where the final spin is measured and the initial spins are aligned. Now we consider an ensemble of unaligned initial spins and the effect of averaging over final spins. We will use the notation $\mathcal{P}_{i;j}(\pb)$ or $\mathcal{P}_{i}(\pb;\ket{j})$ for the probability amplitude given initial spin $j\in [\uparrow, \downarrow]$ and final spin $i\in [\uparrow, \downarrow]$ and momentum $\pb\in\mathbb{R}^3$. For averages over initial, final and both spins, we will use $\mathcal{P}_{i;}(\pb)$, $\mathcal{P}_{;j}(\pb)$, and $\mathcal{P}_{;}(\pb)$, respectively.

Firstly, we will average incoherently over initial spin orientations, for simplicity we will continue with the case of hydrogen. 
In this case, the spin state $\ket{\Phi_{0 \pm 1/2}}=\ket{z^{\pm1/2}_{00}}\ket{\psi_{00}(t')}$, where either $z_{00}^{-1/2} \rightarrow 0$ (spin down) or $z_{00}^{+1/2} \rightarrow \infty$ (spin up). Spatial rotations of the initial state can cover all possible values of the initial $z_{00}$. We can show this explicitly by integrating over the Euler angles
\begin{align}
	\mathcal{P}_{\uparrow;}(\mathbf{p})
        %&=\int d \rho \mathcal{P}_{\uparrow}(\mathbf{p};\mathcal{R}_{\rho}\ket{\uparrow})\\
	&=\frac{1}{8\pi^2}
	\int_{0}^{2\pi}d\alpha\int_{0}^{\pi}d\beta\int_{0}^{2\pi}d \gamma
	\sin(\beta) \mathcal{P}_{\uparrow}(\mathbf{p}; \mathcal{R}_{\alpha \beta \gamma}\ket{\uparrow})
 \label{eq:EulerIntegration1}
\end{align}
This is a well-known result but in the Appendix \ref{sec:Spin-Averaging} we show how to do this using coherent spin states, the result is
\begin{align}
	\mathcal{P}_{\uparrow;}(\pb)=
	\frac{1}{2}\left(\mathcal{P}_{\uparrow; \uparrow}(\pb)+\mathcal{P}_{\uparrow; \downarrow}(\pb)\right).
\end{align}
% Given that, $\mathcal{M}^{\uparrow}_{\mathrm{SO}}(\rb,\pb,t';z_0)\rightarrow a$ as $z_0\rightarrow\infty$ and $\mathcal{M}^{\uparrow}_{\mathrm{SO}}(\rb,\pb,t',z_0)\rightarrow b$ as $z_0\rightarrow0$. 
% \textbf{Is the sentence above something that was used to derive equtaion (33)? What happen to the time argument in $a$ and $b$?}
%
By the same logic $\mathcal{P}_{\downarrow;}(\pb)=\frac{1}{2}(\mathcal{P}_{\downarrow; \uparrow}(\pb)+\mathcal{P}_{\downarrow; \downarrow}(\pb))$.

If we do not measure spin, the positive operator valued measurement (POVM) is given by $\rho=\frac{1}{2}(\ket{\downarrow}\bra{\downarrow}+\ket{\uparrow}\bra{\uparrow})$. This may be applied to a state where the spins are initially prepared to give
$\mathcal{P}_{;\uparrow}=\frac{1}{2}\left(\mathcal{P}_{\uparrow;\uparrow} + \mathcal{P}_{\downarrow;\uparrow}\right)$ and $\mathcal{P}_{;\downarrow}=\frac{1}{2}\left(\mathcal{P}_{\uparrow;\downarrow} + \mathcal{P}_{\downarrow;\downarrow}\right)$.
If we instead apply the POVM to unaligned spins, we obtain the following probability $\mathcal{P}_;(\pb)=\frac{1}{2}(\mathcal{P}_{\uparrow;}(\pb)+\mathcal{P}_{\downarrow;}(\pb))=\frac{1}{4}\left(\mathcal{P}_{\uparrow;\uparrow}+\mathcal{P}_{\uparrow;\downarrow}+\mathcal{P}_{\downarrow;\uparrow}+\mathcal{P}_{\downarrow;\downarrow}\right)$.

In order to isolate the effect of spin, we define a parameter that could feasibly be measured in experiment.
This is $\Delta \mathcal{P}_{;\uparrow_x}(\pb)$, defined as the difference between spin initially aligned in the $x$-direction and unaligned spins given by the difference between spins aligned in the $x$-direction vs fully unaligned spins, denoted $\mathcal{P}_{\uparrow_x}(\pb)$ and $\mathcal{P}_{;}(\pb)$, respectively. In the $z$-basis these probabilities may be written as
\begin{align}
    \mathcal{P}_{;\uparrow_x}(\pb)&=
    \frac{1}{4} \left| 
                                    M_{\uparrow_z;\uparrow_z}(\pb)+M_{\uparrow_z;\downarrow_z}(\pb)
                                    \right|^2 
                                    \notag\\
                                    &\quad+\frac{1}{4} \left|
                                    M_{\downarrow_z;\uparrow_z}(\pb)+M_{\downarrow_z;\downarrow_z}(\pb)
                                    \right|^2
\intertext{and}
    &\mathcal{P}_{;}(\pb)=\notag\\
    &\frac{1}{4}\left(
                |M_{\uparrow_z;\uparrow_z}|^2+|M_{\uparrow_z;\downarrow_z}|^2+|M_{\downarrow_z;\uparrow_z}|^2+|M_{\downarrow_z;\downarrow_z}|^2
    \right).
    \label{eq:P_av}
\end{align}
%We can simplify this by recognising that $M_{\uparrow_z;\uparrow_z}=M_{\downarrow_z;\downarrow_z}^*$ and $M_{\uparrow_z;\downarrow_z}=M_{\downarrow_z;\uparrow_z}^*$.
Now the difference can be expressed as 
\begin{align} 
\Delta \mathcal{P}_{;\uparrow_x}(\pb)&=\mathcal{P}_{;\uparrow_x}(\pb)-\mathcal{P}_{;}(\pb)\notag\\
&=\frac{1}{2}\Re[M_{\uparrow_z;\uparrow_z}(\pb)M_{\uparrow_z;\downarrow_z}^*(\pb)]
\notag\\
&\quad
+\frac{1}{2}\Re[M_{\downarrow_z;\uparrow_z}(\pb)M_{\downarrow_z;\downarrow_z}^*(\pb)].
\label{eq:SpinDiff}
\end{align} 
This provides an `interference' term due to the fact that spin aligned in the $x$-axis but we expressed in terms of the $z$-basis. This provides an observable that is linear in terms of the spin-flip amplitude.
%\textbf{This ends a bit abrupt - any insights related to equations (36) that should be added?}
The difference term in \eqref{eq:SpinDiff} may be written more clearly by expanding the transition amplitudes in terms of the sum over saddle points
\begin{align}
M_{\uparrow_z;\uparrow_z}&=-i\sum_{s}C_{s} a_{s} e^{i S_{s}}\\
M_{\uparrow_z;\downarrow_z}&=-i\sum_{s}C_{s} b_{s} e^{i S_{s}},
\end{align}
where the subscript $s$ refers to the saddle-point solution being summed over, the spin-orbit amplitude $\mathcal{M}_{\SO}$ (\eqref{eq:SpinOrbitTransition}) has been replaced by the coefficient $a_s$ or $b_s$, while $C_s$ is the prefactor defined in \eqref{eq:RelPref}.
The arguments $\pb$, $\rb$, $t$ and $z$ have been dropped from $a_s$, $b_s$ and $C_s$ for simplicity.
Thus, the difference can be determined entirely by products of $a$ and $b$. In the next section, we will solve the equations for $a$ and $b$ analytically to better understand the behaviour of \eqref{eq:SpinDiff}.

\subsection{Analytical Results}
In this section we will consider some analytical results, firstly we will solve the equations of motion for the spin dynamics and secondly derive an approximate analytic approximation to the spin-orbit action.
As we are using the weak approximation in this work, see \secref{sec:RelativisticCQSFA}, the electron trajectories are unaffected by spin-orbit coupling and retain cylindrical symmetry. Furthermore, we restrict our analysis to the $y$--$z$ plane, which means the trajectories will only have angular momentum in the $x$-direction, with $L_z=L_y=0$ 
%\textbf{Does the reader know about the Lz and Ly at this point? - I think we need a little more explanation  here, such that it will be easier to follow}
, the spin-orbit equations in the weak-coupling limit greatly simplify. The coefficients $a(t)$ and $b(t)$, that enter \eqref{eq:SpinOrbitTransition} and parameterize the spin dynamics, see \eqref{eq:ab-ode}, can be found analytically (see Appendix \ref{sec:Ana-Sol} for derivation). The parameter responsible for
spin conserving transitions, is given by
\begin{align}
    a(t)&=\cos(S_{\SO}) \;
    \text{ and the parameter for spin-flips}\notag\\
    b(t)&=-i\sin(S_{\SO}),
\end{align}
where the term $S_{\SO}$ that we will refer to as the spin-orbit action is given by
\begin{align}
    S_{\SO}&=\int_{\Re[t']}^{t} d\tau H_{SO}(\tau)
    \label{eq:SpinOrbit-Action}
\intertext{with}
H_{SO}(\tau)&=\mathbf{C}_{SO}(\rb,\pb)\cdot\mathbf{n}(z)
\label{eq:Hso}
\intertext{and}
\mathbf{C}_{SO}&=\frac{\frac{\partial}{\partial r}V(|\rb|)}{2c^2|\rb|} \Lb.
\label{eq:Cso}
\end{align}
Here $\mathbf{n}(z)=(n_1,n_2,n_3)$, $n_1+i n_2=2 z^*/(1+|z|^2)$, $n_3=-(1-|z|^2)/(1+|z|^2)$ and $\Lb=\rb\times\pb$.
%\textbf{I think the object $H_{SO}(\tau)$ in the equation above needs be clearly defined. This will be important for the possibility for following the steps after equation (40) leading to equations (41)}
%\textbf{I do  not follow the setence above. Maybe better first to refer to an equation for the difference term and then say that it is shown in the figure?(do you really want to refer to fig 6 at this point?), and then say that the difference term can be expressed more clearly}. It is \textbf{The following are not the difference term?}

%\textbf{Explain the meaning of the subscripts on $a$ and $b$ and $C$ in equations (38) and (39). Repeat where C is  defined}
The weak-coupling approximation requires that the spin-obit action $S_{\SO}$ is much smaller than the remaining action $S$. This means that $a$ and $b$ can be expanded in a power series of $S_{\SO}$, such that $a(t)\approx1$ and $b(t)\approx-iS_{\SO}$. Hence, spin-flip probability will be quadratic in $S_{\SO}$ and may be neglected $|M_{\uparrow_z;\downarrow_z}|^2\approx0$ to linear order, which may be expected for the small spin-orbit coupling of hydrogen. 
However, the difference term given by \eqref{eq:SpinDiff} has mixed terms like $a b^*$ and will be of linear order in $S_{\SO}$ and so some effect whereby different alignment of spin affect the final probability distribution is expected.

The spin-orbit action term $S_{\SO}$ can be accurately approximated by assuming contributions only occur when the trajectory is very close to the origin and undergoing Coulomb dominated dynamics, see \figref{fig:RelSurprise}. Thus, during the recollision it is an accurate approximation to neglect the laser field and assume the trajectories follow Kepler hypoerbolae, or when including relativistic corrections to the kinetic energy, the trajectories will follow relativistic corrected Kepler hypoerbolae \cite{lemmon_kepler_2016}, which for a $-Z/r$ potential takes the form
\begin{align}
    \frac{1}{r}=\bar{C}_0\left(
        1+\bar{e}\cos(\bar{\kappa} (\theta-\theta_c))
    \right),
    \label{eq:Relhypoerbolae}
\end{align}
where $\bar{C}_0=(1+\frac{1}{2}\epsilon)(Z/l^2)$, $\bar{\kappa}=1-\frac{1}{2}\epsilon$, with the corrected eccentricity $\bar{e}=(1+\frac{1}{2} \epsilon)e$ written in terms of the Kepler eccentricity $e=\sqrt{1+2E l^2/Z^2}$, with $E$ being the energy and $l$ the magnitude of the angular momentum in the non-relativistic case. The relativistic correction parameter $\epsilon=1/(\bar{l}^2c^2)$, where the magnitude of the relativistic angular momentum $\bar{l}\approx l(1+v^2/c^2)$. In the case without relativistic corrections to the kinetic energy $\epsilon=0$.
%\textbf{I think the following instruction leading to equation (41) needs to be explained a little more -this comment relates to my comment after equation (37)}
The spin-orbit action $S_{\SO}$, given in \eqref{eq:SpinOrbit-Action}, may be written more explicitly as
\begin{align}
    S_{\SO}&=\frac{Z}{2c^2}
    \int_{\Re[t']}^{t}
    \frac{\bar{l}d\tau }{r^3},
\end{align}
by substituting the integration variable from time $t$ to the orbital angle $\theta$ via $r^2 d \theta = \bar{l} d\tau$ and substituting in \eqref{eq:Relhypoerbolae} the spin-orbit action can be approximated as
\begin{align}
    S_{\SO}\approx \frac{\bar{C}_0}{2 c^2}\left(
        \arccos\left(-\frac{1}{\bar{e}}\right)
        +\frac{\bar{e}}{\bar{\kappa}}\sqrt{1-\frac{1}{\bar{e}^2}}
    \right).
    \label{eq:SpinOrbit-Analytic}
\end{align}

This equation is tested in \figref{fig:RelSurprise}, using the position and velocity when the recolliding trajectory first reaches a distance of $1$~a.u. from the residual ion, the value of the spin-orbit action phase, $S_{\SO}$,  is computed and placed on the figure as the circles in panels (c) and (f). %\textbf{Where should I look in the figure - add instruction and modify caption? }. 
This works exceptionally well for the non-relativisitic case [\figref{fig:RelSurprise}(c)], with no difference visible between it [coloured circle] and the numerical value [dot-dashed line]. It also works well for the case including relativistic correction to the kinetic energy [\figref{fig:RelSurprise}(f)], however, the approximated value (represented by a circle) is slightly off the numerical value (plotted as a dot-dashed line), indicating the analytical estimation slightly worsens in the relativistic case.
% \textbf{Add information about where the reader should look in \figref{fig:RelSurprise}. I guess some information about this approximative result needs to be added to the caption of \figref{fig:RelSurprise}}.
%\textbf{I assume this is because the constants in \eqref{eq:Relhypoerbolae} are not actually constant in the relativistic expansion, but I have to assume they are approximately constants for the sake of the integration. But I have to verify this.}

\begin{figure*}
    \centering
    \includegraphics[width=0.65\linewidth]{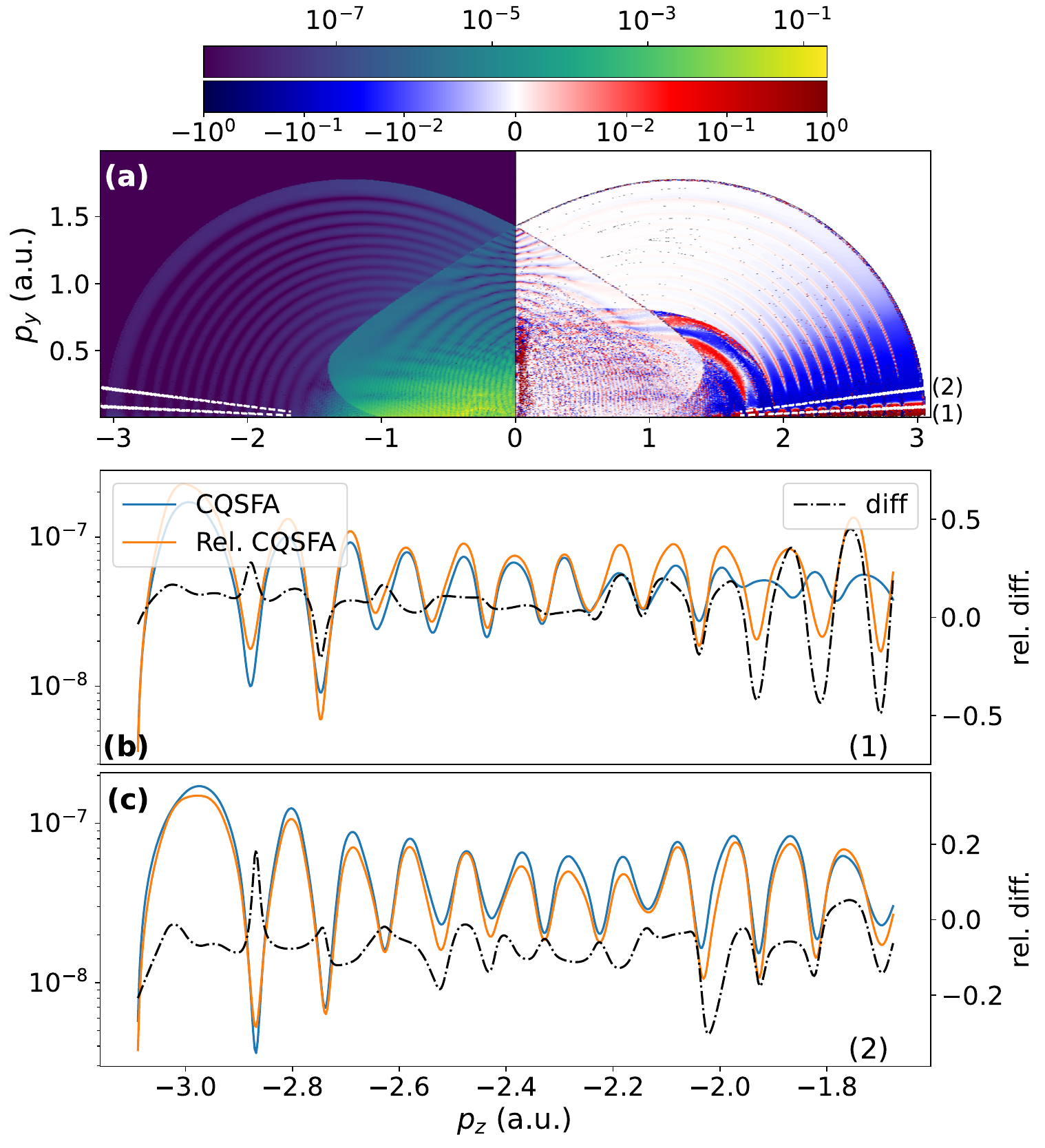}
    \caption{The difference between momentum distributions with and without relativistic corrections. Spin and spin-orbit effect have been neglected by setting the $\mathcal{M}_{SO}=1$ in \eqref{eq:RelTransSaddle}.
    (a) left is the relativistic momentum distribution for a hydrogen target, averaged over initial and final spins, $\mathcal{P}_{;}(\pb)$ of \eqref{eq:P_av}. (a) right is the difference between this and the distribution without relativistic corrections to the kinetic energy. %\textbf{Is spin included somewhere? - the caption should be precise about this issue. The main text indicates that spin-orbit is absent in this figure. }.
    (b) and (c) include a lineout of (a) left [solid line left] and (a) right [dashed right], given by the lower [labelled (1)] and upper [labelled (2)] white lines in (a), respectively. %\textbf{the labels are only to the left. The previous sentence tells me that I should look at lines both in the left and right parts of figure 5(a)}
    }
    \label{fig:RelDiff}
\end{figure*}

\begin{figure*}
    \centering
    \includegraphics[width=0.65\linewidth]{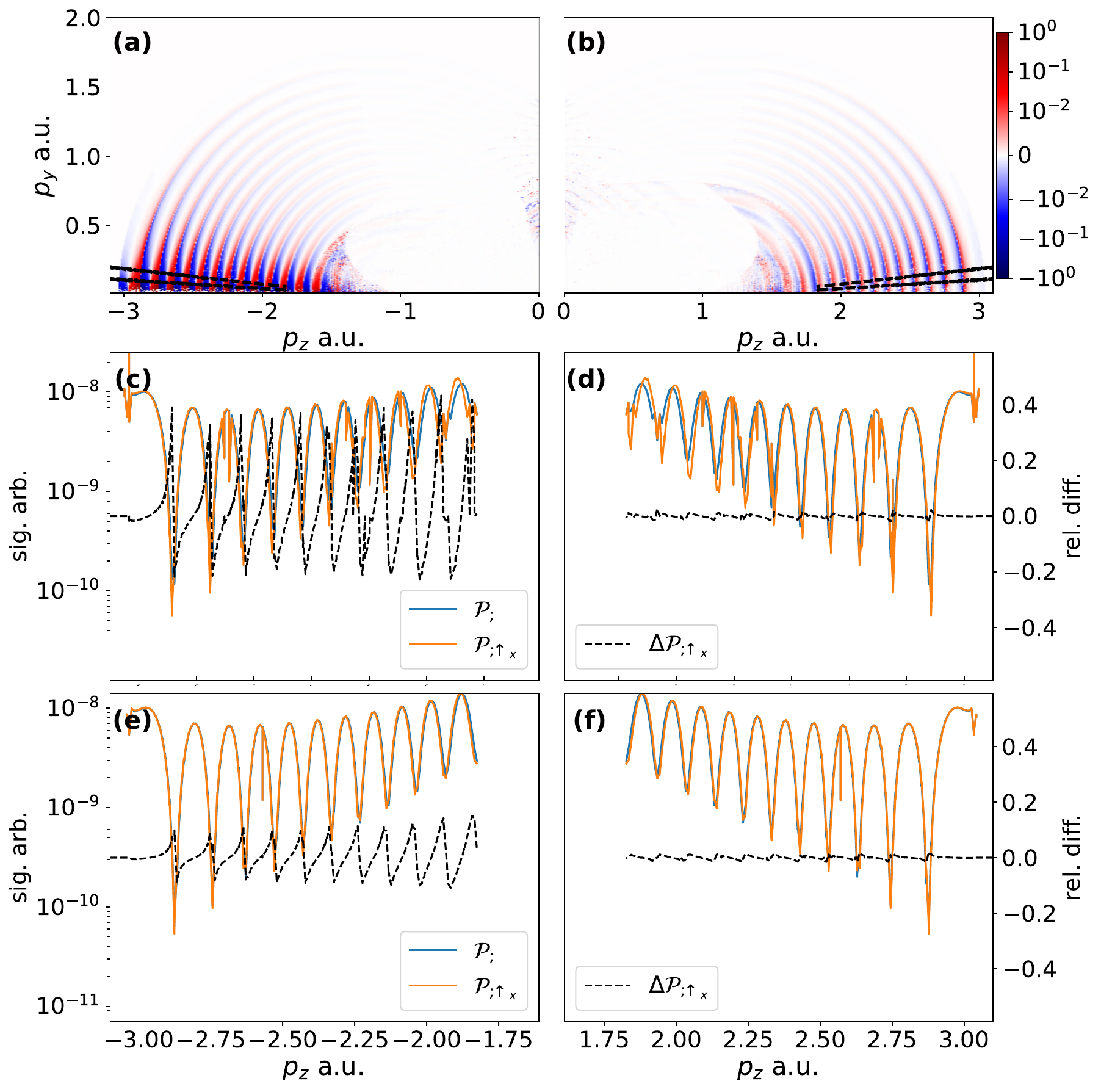}
    \caption{The difference between momentum distributions with initial spin aligned in the $x$-direction and totally averaged spins, given by $\Delta \mathcal{P}_{;\uparrow_x}(\pb)$ defined in \eqref{eq:SpinDiff}. The left column [(a), (c), (e)] neglects the corrections the kinetic energy, the right column [(b), (d), (e)] includes corrections to the kinetic energy.    
    (a) and (b) show the relative difference over the full momentum region. The lineouts, given by the black lines in (a) and (b), show the signal of averaged spins $\mathcal{P}_{;}(\pb)$ and $x$-aligned spins $\mathcal{P}_{;\uparrow_x}(\pb)$, and relative difference $\Delta \mathcal{P}_{;\uparrow_x}(\pb)$. As in \figref{fig:RelDiff}, the lower lineout in (a) [(b)] corresponds to (c) [d], while the upper lineout corresponds to (e) [(f)].}
    \label{fig:SpinDiff_NoRel}
\end{figure*}

\section{Relativistic Results}
\label{sec:RelResults}
The trajectories involved in the relativistic computation are shown in the bottom row of \figref{fig:RelSurprise}. With relativistic corrections to the kinetic energy, the velocity of the rescattered trajectories, \figref{fig:RelSurprise}(d), no longer exceeds the speed of light. There is also a directly related effect on the minimum distance to the core, \figref{fig:RelSurprise}(e), which is reduced by more than two orders of magnitude compared to the  non-relativistic case in \figref{fig:RelSurprise} (b).
We can gain insight into why the core is probed less strongly when relativistic corrections are included, by examining the problem from a non-relativistic perspective. We may imagine that there is an effective `repulsion' (dependent on the velocity), which balances the attraction of the Coulomb potential, meaning that the electron does not get so close to the core, only reaching around $10^{-3}$~a.u.
This effect can be seen clearly by rearranging the saddle point equations of motion (neglecting the laser field) to give
\begin{equation}
    \ddot{\rb}=(\id_3-\frac{1}{2c^2} \underline{\underline{M}} )\mathbf{F},
    \label{eq:Mod_force}
\end{equation}
where $\mathbf{F}=-\nabla V(\rb)$ is the classical non-relativistic force due to the potential and $\underline{\underline{M}}$ is a matrix given by $\underline{\underline{M}}=\id_3 \pb^2 + 2\pb \otimes\pb$. For a derivation of \eqref{eq:Mod_force}, see Appendix \ref{sec:classical-corrections}.
Now it is possible to see, following Appendix \ref{sec:classical-corrections}, that for a trajectory with momentum $\pb$ confined to the $z$-axis, the attractive force of the potential is reduced by the factor $1-\frac{3}{2}\frac{v^2}{c^2}$, while for circular motion with momentum perpendicular to the origin this factor becomes $1-\frac{1}{2}\frac{v^2}{c^2}$.
In both cases, this leads to trajectories that do not probe the core as closely as in the non-relativistic case.
The result of the significant increase in the minimal distance from the core and the reduction in the velocity is that the spin-orbit action phase term is greatly reduced, by an order of magnitude for the directly recolliding trajectory and by over a factor of 5 for the long and short trajectory pairs, and the phase for all trajectories are over an order of magnitude below the phase due to the Coulomb potential. 
Thus, without relativistic corrections to the kinetic energy, the spin-orbit coupling phase is overestimated by around an order of magnitude, while with relativistic corrections, spin-orbit coupling is quite modest. The relatively small action phase of the spin-orbit coupling, in this case, also validates the weak-coupling approach used in the relativistic CQSFA of \secref{sec:RelativisticCQSFA}.

Another important observation that was already discussed in \ref{sec:Analytic} is that the analytical approximation for the spin-orbit phase, given by \eqref{eq:SpinOrbit-Analytic}, provides a very good approximation when compared to the numerical value, in FIGS.~\ref{fig:RelSurprise}(c) and (f), for the case without and with the kinetic energy corrections, respectively. 
%This can save computation time 
%In fact, an alternative approach such as using the saddle point approximation on the spin-orbit path-integral may be a worse approximation in this regime.

% \begin{figure*}[t]
%     \centering
%     \includegraphics[width=0.65\linewidth]{Figures/SpinDiff_Mono_1Cycle_H_1600_0.5e14.pdf}
%     \caption{The same a \figref{fig:SpinDiff_NonRel} but computed including the relativistic corrections to the kinetic energy. }
%     \label{fig:SpinDiff_Rel}
% \end{figure*}

In \figref{fig:RelDiff}, we show the PMD for a linearly polarized monochromatic laser field, where ionization is allowed only during a single cycle. The target is hydrogen as before and the intensity and wavelength are again $5\times10^{13}$~W/cm$^2$ and 1600 nm, respectively. The consideration of a single cycle provides a simple PMD, with fewer trajectories for an easier analysis, while also maximizing the rescattered trajectories' return energy, thus, showing the largest relativistic effects. Similar results can still be seen for relativistic computations using a sin$^2$ envelope (not shown). In this figure, we have neglected the spin and spin-orbit effect to isolate the kinetic energy corrections, $K_1(\pbh)$ in \eqref{eq:K-Corr}. 
%\textbf{Probably this information about absence of spin should be given also in the caption of the figure}
The left side of \figref{fig:RelDiff}(a) plots the PMD including relativistic corrections to the kinetic energy, where the same high-energy rings $(p_{fz}, p_{fy})\approx(-3.0, 0.1)$~a.u. can be seen as in \figref{fig:TDSE_CQSFA}(a). The right side of \figref{fig:RelDiff}(b), shows the normalized difference between the CQSFA with and without relativistic kinetic energy corrections.
We find that in the high-energy rescattering region near the $p_z$-axis a series of red peaks occur, which shows that the CQSFA PMD signal with relativistic kinetic energy corrections is larger than without. However, away from the $p_z$-axis this situation reverses.
In FIGS.~\ref{fig:RelDiff}(b) and (c), we look at these regions in more detail by plotting the lineouts, marked on panel (a).
The first lineout, near the axis (b), clearly shows that the relativistic case exceeds the non-relativistic case, with the normalized difference exceeding 50\%. While the situation is reversed in \figref{fig:RelDiff}(c) and the normalized difference approaches $20\%$.

The main driver of this effect is an overall change in probability for the pair of rescattered trajectories that contribute to this region.
This is because scattering with and without relativistic corrections only appreciably changes $\pb_{0\perp}$, the initial momentum perpendicular to the laser polarization direction, which leaves the ionization times and tunnelling probability almost the same with and without relativistic corrections to the kinetic energy. 
The case with relativistic corrections leads to a larger value of initial perpendicular momentum $|\pb_{0\perp}|$ than the non-relativistic case. The perpendicular momentum controls how close the electron trajectory gets to the ion, with a higher value meaning the core gets probed less strongly in the case with relativistic corrections.

These changes in initial conditions affect the so-called stability factor $1/\sqrt{|J|}$, determined from the Jacobian by $J=\frac{\delta \pb_f}{\delta \pb_0}$ in \eqref{eq:TransSaddle}, see also \eqref{eq:RelPref}, which leads to the visible changes in the PMDs. In the case of backscattering, where the scattering angle is nearly $180^\circ$, the trajectories for the case including kinetic energy correction undergo a much smaller momentum change than those without the kinetic energy corrections. This means they are less sensitive to the initial conditions. On the other hand, for scattering angles less than approximately $175^\circ$, the trajectories including relativistic corrections have a comparable momentum transfer to those without the corrections. However, relativistic scattering can lead to a greater scattering angle than the completely non-relativistic case, meaning now the trajectories including relativistic corrections to the kinetic energy are more sensitive to the initial conditions than the case without corrections.
Now we will consider the effect of spin-orbit coupling. It is important to note that in this article we only consider the final electron momentum in the $p_{fz}$-$p_{fy}$ plane, and the equations of motion for the electron are cylindrically symmetric \footnote{This is because, in the weak-coupling limit, the effect of spin-orbit coupling on the electronic motion is neglected.}.
Thus, the trajectories are restricted to the $zy$-plane, and the angular momentum is in the $x$-direction and couples to spin in this direction. For motion in rotated planes, e.g., the $p_z$-$p_x$ plane the angular momentum and spin-orbit coupling will be rotated accordingly.
%So in all following results we will consider the effect

In \figref{fig:SpinDiff_NoRel}, the effect of spin-orbit coupling is considered on the initial alignment of the electronic spin. \figref{fig:SpinDiff_NoRel}(a) shows the relative difference between electronic spin aligned in the positive $x$-direction vs unaligned spins, where relativistic kinetic energy corrections have been neglected. 
This combination of spin alignment is considered, as it could be feasibly done in experiment and measured, using a B-field for spin alignment.
Lineouts near the $p_z$-axis are shown in panels (c) and (e) (the same as in \figref{fig:RelDiff}, where the largest effect, due to the backscattered trajectories, can be observed). 
Here, the relative difference approaches 50\%, this is surprising given the small degree of spin-orbit coupling expected for a hydrogenic target, and this difference could foreseeably be measured. 
%\textbf{This effect is driven by a small but non-zero spin flip component. I found this a bit confusing at first, as I thought due to conservation we couldn't get a spin flip for the Coulomb system.
%However, on closer inspection it seems like the systems of equations in the weak-coupling approximation do predict this, see below. I think in the weak approximation, the spin-flip parameter $b$ can change but should remain small enough such that $b^2\approx0$, which is more or less the case. This means we can observe effects due to different alignment through mixed terms $a b^*$ but spin-flips are still effectively zero probability.}

As argued above, a correct treatment requires that corrections to the kinetic energy be taken into account, so we plot the same in \figref{fig:SpinDiff_NoRel}(b), (d), and (f) but include these corrections. This leads to much weaker differences,  demonstrating the importance of including the relativistic kinetic energy correction for the rescattered electron when considering spin-orbit coupling. 
The difference with and without the relativistic kinetic energy corrections can be explained by the behaviour of in the spin-orbit phase, which is far too big if kinetic energy corrections are not taken into account, see \figref{fig:RelSurprise}. 

\begin{figure*}
    \centering
    \includegraphics[width=\linewidth]{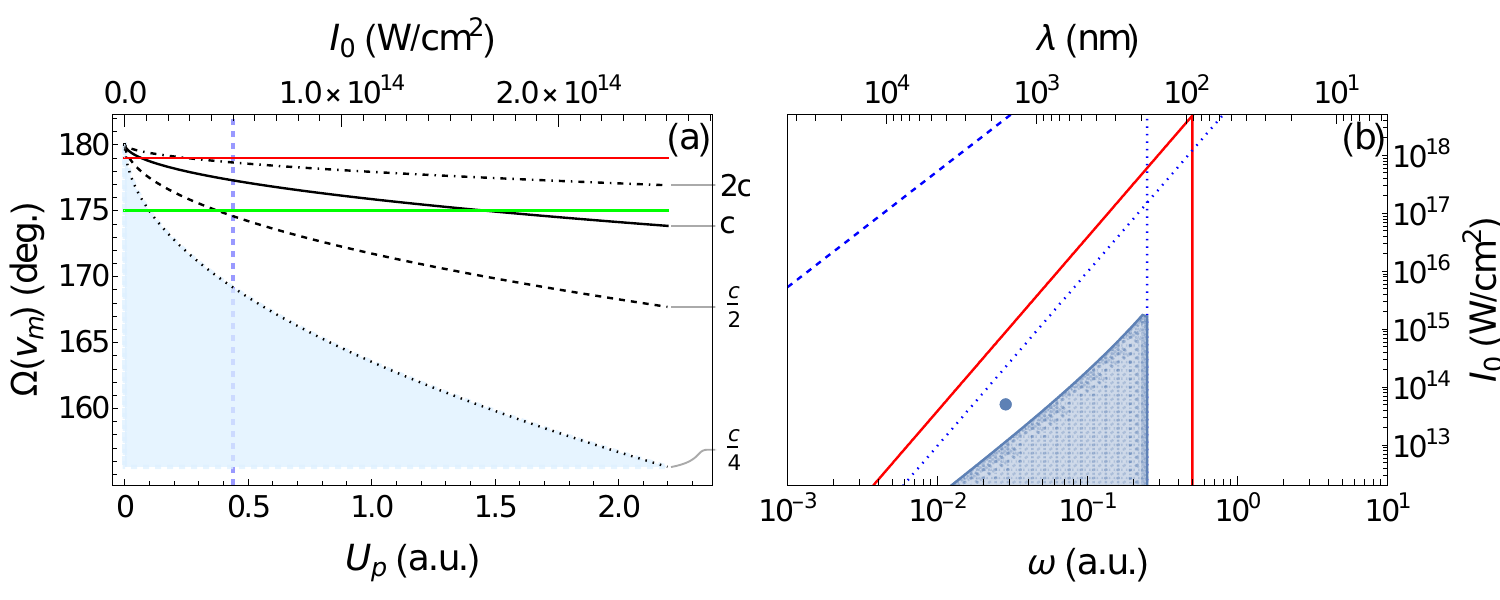}
    \caption{The limits of a non-relativistic theory for rescattered photoelectrons. (a) the scattering angle, $\Omega(v_m)$, for particular maximum rescattering velocities ($v_m=c/4$, $c/2$, $c$, and $2c$, given by the dotted, dashed, solid, and dot-dashed black curves, respectively) across a range of intensities, as given by \eqref{eq:limits_new}. The shaded region below the curve for $v_m=c/4$ is the region where relativistic effects will not play a role. The wavelength (angular frequency) is fixed to $1600$~nm ($0.0285$~a.u.), the red and green horizontal lines denote a $179^\circ$ and $175^\circ$ scattering angle, respectively. The dashed blue vertical line denotes the intensity of $I_0=5\times10^{13}$~W/cm$^2$ ($\up=0.44$~a.u.), used in this study. (b) for a range of angular frequencies and intensities, the region where a non-relativistic theory holds for the rescattered photoelectrons is given by the blue shaded region, as given by \eqref{eq:limits_new}. This assumes a maximum scattering velocity of $0.25 c$ and a maximum scattering angle of $175^\circ$. The blue dashed line denotes the typical relativistic condition, $I_0=2 c^2 \omega^2$, while the red solids line denoted the dipole conditions, $\omega=1/2$ and $I_0=8 c \omega^3$, outside of which the dipole approximation breaks down. Similarly, the blue dotted lines marked the region where a tunnelling model will hold, $\omega=1/4$ and $I_0=2 c \omega^3$, as given in \cite{reiss_limits_2008}. A blue dot is used to show the parameters used in this study. }
    \label{fig:NewLimits}
\end{figure*}

\section{New Limit on Non-Relativistic Theories}
\label{sec:NewLimits}
The results from \secref{sec:RelResults} highlight the importance of relativistic corrections for the rescattered portion of the wavepacket in strong-field ionization. However, commonly employed estimations for when relativistic effects should be accounted for in the infrared regime rely on classical arguments considering the motion of the free electron in the continuum \cite{protopapas_atomic_1997, reiss_limits_2008} and do not account for scattering. In this section, limits will be derived that account for the scattering behaviour and we will compare to existing limits. 

A commonly employed limit for relativistic behaviour is when the ponderomotive energy approaches the electron rest mass energy, that is $\up=m c^2/2$ \cite{reiss_limits_2008}, this leads to a condition on the intensity $I_0=2 c^2 \omega^2$ above which relativistic effects are significant. This is plotted as blue dashed line in the upper left corner of \figref{fig:NewLimits}(b), the blue dot indicates the parameters used here are well below this intensity. Another important condition is the region where the dipole approximation is valid, this is given by an `upper' and `lower' limit. The `upper' limit is simply that the angular frequency needs to be small enough and is given by $\omega=1/2$ \cite{reiss_limits_2008}. The `lower' limit derives from the requirement that the motion of the free electron due to the laser magnetic field should be small and leads to the following limit on intensity $I_0=8 c \omega^3$, where higher intensities mean the break-down of the dipole approximation. These upper and lower limits are given by the red solids lines in \figref{fig:NewLimits}(b). A slightly smaller region, where tunnelling models will hold, is also given by the blue dotted lines  in \figref{fig:NewLimits}(b), see \cite{reiss_limits_2008}.

To derive a new condition, we must consider the maximum velocity that a rescattering trajectory will achieve given specific laser parameters and particular scattering angle. We will consider angles close to backscattered, as these represent the most extreme cases. The most energetic rescattered trajectories gain around $3.17\up$ kinetic energy from the laser field upon return. If we assume once the electron is close enough, $r_\mathrm{B}$ from the core, it undergoes Coulomb dominated dynamics, and we may ignore the field from this point, then an energy conservation argument may be used to approximate the maximum velocity. The total energy entering this boundary may be approximated as $E_{\mathrm{B}}=3.17 \up -Z/r_0$, where the potential energy convert to kinetic energy from the tunnel exit at $r_0$, to the boundary ($r_\mathrm{B}$) has also been accounted for. Equating $E_{\mathrm{B}}$ with the energy at closest approach $r_m$, gives an equation for the maximum velocity
\begin{align}
    \frac{1}{2}v_{m}^2&=E_{\mathrm{B}}+\frac{Z}{r_m}
    \intertext{assuming Coulomb dominated dynamics we may substute the equation for a Kepler hyperbola at the closest approach, see \eqref{eq:Relhypoerbolae},}
    \frac{1}{2}v_{m}^2&=E_{\mathrm{B}}+\frac{Z^2}{l^2}
    \left(
        1+e
    \right),
    \intertext{where for nearly backscattered trajectories $e\approx1$ }
    \frac{1}{2}v_{m}^2&=E_{\mathrm{B}}+\frac{2 Z^2}{l^2}.
\end{align}
Now the angular momentum may be written in terms of the maximal velocity
\begin{equation}
    l^2 = \frac{4 Z^2}{V_m^2-2 E_{\mathrm{B}}}.
    \label{eq:limits_l}
\end{equation}
The dynamics will depend on the scattering angle, the closer to backscattering the higher the maximum velocity will be. Thus, we can use the scattering angle to help formulate the condition. The scattering angle $\Omega=2\Psi-\pi$, where $\Psi$ is the asymptote angle, that is the angle between the direction of the incoming asymptotic and closest approach (periapsis) direction, which is given by
\begin{align}
    \cos(\Psi)=-\frac{1}{\sqrt{1+\frac{2E_{\mathrm{B}} l^2}{Z^2}}}.
    \intertext{Inserting \eqref{eq:limits_l} gives an equation for the scattering angle in terms of the maximal velocity}
    \Omega(v_m)=2\arccos \left[
    -\frac{1}{\sqrt{1+\frac{8 E_{\mathrm{B}} }{v_m^2-2 E_{\mathrm{B}}}}}
    \right]-\pi.
    \label{eq:limits_new}
\end{align}
 In \figref{fig:NewLimits}(a), contours of \eqref{eq:limits_new} are given for different values of $v_m$ in terms of $c$. For an increasing laser intensity, the scattering angle (given a fixed maximal velocity) reduces. For the curve, $\Omega(c/4)$, this is used to denote a region (shaded in blue), where relativistic effects may be neglected. As intensity increases, this region reduces so that less of the rescattered wavepacket may be accurately described by a non-relativistic model, as higher velocities will be reaches during rescattering. The parameter region in this study (blue dashed line), crosses this boundary for scattering angle greater than $170^\circ$, and crossed the $\Omega(2 c)$ at around $179^\circ$. This is confirmed by \figref{fig:RelSurprise}, where the long and short orbit 4 trajectories scattered by $179^\circ$ reach velocities just over $2c$, when neglecting corrections to the kinetic energy. Note that orbit 4b would actually require a strong condition, as it has a return energy of higher than $E_{\mathrm{B}}$ and returns with a $6c$ velocity. However, for these parameters, this solution has low probability.
 
 In \figref{fig:NewLimits}(b), we use this condition to construct a modified region (blue shaded region), inside the dipole allowed region, where relativistic effects are not important for scattering angles $\Omega(0.25 c)<175^\circ$, given a maximal velocity of $v_m=0.25c$. This reduces the standard dipole allowed region by around a factor of 2, and the blue point representing the parameters used in this study lies outside this region despite being inside the dipole allowed region. Note, the typical relativistic condition (blue dashed line in the upper left of the figure) lies more than 6 order of magnitude in intensity from our modified condition, demonstrating the huge difference that rescattering can make when considering relativistic effects.

\section{Conclusion and outlook}
\label{sec:Conclusions}
%\textbf{points we want to make (not necessarily in this order): (i) coherent spin formalism makes inclusion of spin possible (ii) relativistic effects observed in a regime where is is not expected. (iii) Boundaries on the backscattering region within relativistic effects need be onsidered (iv) one generally gets  the S-O coupling wrong without relativistic corrections to the kinetic energy. Maybe some of the following sentences can be used.}
Driving atomic and molecular systems with near-infrared light at intensities up to around 10$^{14}$ W/cm$^2$ was thought to be well described by non-relativistic quantum mechanics \cite{reiss_limits_2008}, besides effects associated with large spin-orbit energy splitting in the cationic system \cite{carlstrom_control_2023}.
In this study, we have used the powerful machinery of the path integral-based CQSFA to investigate spin-orbit coupling and other relativistic dynamics in more detail. For the parameter range explored, we find the CQSFA provides an exceptional quantitative agreement with the single-active electron TDSE. However, upon close inspection, we have illustrated that the non-relativistic treatment breaks down for a significant portion of the rescattered electron wavepacket, at a laser intensity orders of magnitude below that expected.

%\textbf{Maybe in the following 'backscattering' should be changed to 'trajectories that probe the core strongly'}
%The present study confirms this believe for dynamics not involving backscattering of the  laser-driven recolliding electron. For backscattered electrons, however, we have illustrated that the nonrelativistic treatment breaks down. 
The origin of this breakdown is that the backscattered electrons probe the core very closely and gain significant kinetic energy.  In the non-relativistic treatment, this results in unphysical superluminal velocities  of the electron. To remedy this situation, we extended the CQSFA to include all relativistic correction terms from the Briet-Pauli Hamiltonian, including spin-orbit coupling and corrections to the kinetic energy. 
With this improved model we evaluate the effects of these terms, and find the kinetic energy corrections to be significant for the rescattered part of the electron, while the spin-orbit coupling is massively overestimated if computed without the former. 
Accordingly, the assessment of dynamical spin-orbit effects by theory not accounting for relativistic kinetic energy corrections (which is common given its inclusion is numerically intensive), need to take care, since such theory would significantly overestimate the spin-orbit interaction effects. In order to inform future work, we provide an expression for where relativistic effects become important, and which lies many orders of magnitude below the typical limit for relativistic effects. 

The rescattered region in question is important for application in attosecond physics and chemistry, through imaging processes, such as LIED, where the diffraction pattern of the rescattered electron is used to infer sub-femtosecond nuclear motion in molecules \cite{blaga_imaging_2012,wolter_ultrafast_2016}.
Clearly, an interpretation of an experimental PMD with a model involving non-relativistic cross-sections would lack the spin-orbit and kinetic energy corrections of focus in this work, and could therefore lead, e.g., to deviations between inferred and actual time-resolved distances. Thus, the models used for LIED should incorporate relativistic scattering cross-sections. 
Work assessing quantitatively the implications of our findings for this kind of investigation is currently in progress, including adapting the CQSFA for LIED.

%\textbf{some of this next piece of text is perhaps not in paper-style , maybe more 'cover-letter' style - but the point we want to make in the conclusion in some manner}
%One could perhaps argue that the present findings are mainly of academic interest since the backscattering cross-section is very small compared to the cross-section in other directions [MottandMasseyBook]. In other words, the effect we report is in a regime where there is hardly any signal, so why should we bother? The answer is that the backscattering region is a key region in the PMD for application in attosecond physics and chemistry. It is so because this region is of interest  for techniques such as LIED. In LIED the diffraction pattern of the backscattered electron is used to infer sub-femtosecond nuclear motion in molecules  [BiegertScience2016,BlagaFirstauthorNatureDiMauroGroup]. Clearly, an interpretation of an experimental PMD with a model involving nonrelativistic cross-sections would lack the spin-orbit and kinetic energy corrections of focus in this work, and could therefore lead, e.g., to deviations between inferred and actual time-resolved distances. Work assessing quantitatively the implications of our findings for this kind of investigation is currently in progress.    

The conclusions and perspectives discussed above were obtained from the new theory developments of this work, where an improved non-relativistic quantum trajectory-based CQSFA was used to reach unprecedented agreement with TDSE simulations for PMDs. This agreement allowed us to proceed with a relativistic extension of the CQSFA. The coherent spin formalism and weak-coupling limits allowed such development, including analytical elucidation of the dynamics. The trajectory-based approach benefits from ease of interpretation in terms of possibly interfering quantum paths. Furthermore, the effects of the different coupling terms in the action can easily be isolated. This separation into individual contributions was used here to show the reduction in the spin-orbit action phase by the corrections to the relativistic energy. Such methodologies could also be extended to describe the spin of a residual ion. 
%Which in the future could be coupled with appropriate molecular pseudo potentials to govern dynamics and rescattering. 
Furthermore, the coherent spin state formalism opens up the possibility of including other degrees of freedom supported by this description, such as the quantum state of the laser. Which has recently been found to lead to a broad range of significant effects \cite{gorlach_quantumoptical_2020,lewenstein_generation_2021,stammer_high_2022,rivera-dean_lightmatter_2022,stammer_quantum_2023,pizzi_light_2023,eventzur_photonstatistics_2023}. 
%One approach, connecting the semi-classical equations of motion to the quantum state of the light, has been explored \cite{eventzur_photonstatistics_2023}, but this neglected the Coulomb potential. An approach using the CQSFA would include this along with many other effects such as those considered here.

Looking into the future, it would be very interesting to connect this work with recent work on orbital angular momentum in strong-field ionization \cite{maxwell_manipulating_2020,kang_conservation_2021,planas_ultrafast_2022} and examine the role of total angular momentum in imaging, as well as the implications for inelastic double ionization processes \cite{maxwell_entanglement_2022} demonstrating angular momentum entanglement.
Another important direction, is to extend the present studies to longer mid-infrared wavelengths, say up to around $3000$~nm. This is the wavelength regime of current LIED experiments \cite{blaga_imaging_2012,wolter_ultrafast_2016}. The present CQSFA with relativistic corrections is scalable to this wavelength domain, while TDSE-based approaches would be severely challenged by numerical difficulties related to the increase in the required number of total angular momenta, along with relativistic corrections. In this longer wavelength regime, nondipole corrections to the laser-matter interaction would have to be included \cite{reiss_limits_2008,jensen_nondipole_2020}, and this will affect the initial conditions for the trajectories \cite{madsen_nondipole_2022}, as well as the shape of the PMD and ATI \cite{brennecke_highorder_2018,madsen_disappearance_2022}.

In addition to the scaling with wavelength, our results also show that the interval of backscattering angles, where relativistic corrections are needed, increases with laser intensity.  Mid-infrared intense femtosecond laser pulses are currently becoming more readily accessible, e.g. \cite{wolter_strongfield_2015}, so we envision increased need for theory incorporating relativistic effects in the future. If the rescattering energy becomes sufficiently large, one could even contemplate laser-induced time-resolved investigations of the structure of the atomic nucleus, with an achievable timescale of less than an attosecond made possible by the very brief electron transit, opening up a new domain on sub-attosecond physics.
%Such effects would probably require energies beyond the applicability range of the present approach - but probably some QRS type model could be established and then one could insert the correct crosse section.

%\textbf{Problems with references: [14]: Title incomplete? [23]: write out JOSA? [32]: remove the dollar signs around R [72,75] Add CITY to the information [81]: wrong authorlist}

\begin{acknowledgments}
ASM acknowledges funding support from the European Union’s Horizon 2020 research and innovation programme under the Marie Sk\l odowska-Curie grant agreement SSFI No.\ 887153. LBM acknowledges support from the Danish Council for Independent Research (Grant Nos.\ 9040-00001B and 1026-00040B).
\end{acknowledgments}

\appendix
% \begin{figure*}
%     \centering
%     \includegraphics[width=\linewidth]{Figures/HistogramNoRel_Sin2_6Cycle_H_1600_0.5e14_Sin2_6Cycle_H_1600_0.5e14.pdf}
%     \caption{Histogram of the maximum velocity (left) and minimum distance (right) from the origin for all orbits for with (lower) and without (upper) relativistic corrections to the kinetic energy, for the same six cycle pulse and parameters as \figref{fig:TDSE_CQSFA}. }
%     \label{fig:Hist}
% \end{figure*}
\section{CQSFA theory}
\label{Sec:AppendixCQSFA}
In this Appendix, we discuss some additional approximations that are used in the CQSFA theory.
%\subsection{Additional analysis}

We use the analytical Kepler expression to extend the trajectories and action asymptotically to infinite time, once the laser pulse is over \cite{shvetsov-shilovski_semiclassical_2016,carlsen_precise_2023}.
The tunnel integral over the binding potential is approximated by the Coulomb-factor \cite{perelomov_ionization_1966,bisgaard_tunneling_2004},
\begin{equation}
\int_{t_s}^{\Re[t_s]}V[\rb_s(\tau)]d \tau \approx
\left(
    \frac{4\ip}{|\Eb(t_s)|}
\right)^{\frac{Z}{\sqrt{2 \ip}}}.
\label{eq:CoulombTunnel}
\end{equation}
This approach has two advantages, it automatically regularizes the divergent integral, given by direct evaluation of the integral in \eqref{eq:CoulombTunnel}, and avoids branch cuts that arise in this integral, due to taking the square root of the complex-valued position vector over the tunnel exit. As in previous works \cite{maxwell_coulombcorrected_2017,maxwell_analytic_2018,maxwell_coulombfree_2018}, we take the position to be real for real-time propagation, i.e., $\rb_s(\Re[t_s])=\Re[\rb_0]$ and keep the momentum fixed during tunnelling, i.e., $\pb_s(t)=\pb_{0s}$ for $t\in[t_s, \Re[t_s]]$. Here, the tunnel exit is given by
\begin{equation}
    \rb_0=\Re\left[\int_{t_s}^{\Re[t_s]}d\tau (\pb_{0s}+\Ab(\tau))\right].
\end{equation}
Aside from these approximations specifying the initial conditions for the trajectory propagation, all other parts of the action, given by \eqref{eq:Action}, are computed in full.

\section{\label{sec:SpinStateFrame}Spin-state framework}
In order to describe spin in a path-integral framework, we require coherent spin-states \cite{kochetov_su_1995,pletyukhov_semiclassical_2002} (an irreducible representation for $SU(2)$). In its most general form a $SU(2)$ coherent spin state can be written as
\begin{equation}
	\ket{z;S}=\frac{e^{z S_{+}}\ket{S, m_s=-S}}{(1+|z|^2)^S}.
\end{equation}
For a spin-1/2 ($S=1/2$) system, dropping all the $S$'s, this may be written as
\begin{equation}
 \ket{z}=\frac{\ket{\downarrow}+z\ket{\uparrow}}{\sqrt{1+|z|^2}},
\end{equation}
where we use $\ket{\uparrow}$ ($\ket{\downarrow}$) to denote $m_s=\frac{1}{2}$ ($m_s=-\frac{1}{2}$).
A vital step when deriving a path integral representation of a propagator is insertion of the resolution of the identity in time sliced amplitudes. For coherent spin-states the resolution of the identity may be written as 
\begin{align}
	I_S=\int d\mu_S(z) \ket{z;S}\bra{z;S}, && d\mu_S(z)=\frac{2S+1}{\pi}\frac{d^2 Z}{(1+|z|^2)^2}
\end{align}

The coherent spin states map spin states to the complex plane, see \figref{fig:CoherentSpin}. Some key values are $\ket{z\rightarrow0}=\ket{\downarrow}$, $\ket{z\rightarrow\infty}=\ket{\uparrow}$, $\ket{z\rightarrow1}=\frac{1}{2}(\ket{\uparrow}+\ket{\downarrow})$ and $\ket{z\rightarrow-1}=\frac{1}{2}(\ket{\uparrow}-\ket{\downarrow})$.

%\section{\label{sec:BuildSpinStates}Building spin states}
It is useful to apply this formalism to building eigenstates of the field-free system, i.e. $\hat{H}_{0,\SO}=\hat{H}_0+\hat{H}_{\SO}$. Consider an initial state with quantum numbers $j$ and $m_j$, built using the standard angular momentum addition rules, e.g., see Ref.~\cite{barth_spinpolarized_2013}.
\begin{equation}
	\ket{\Phi_{j m_j}}=\sum_{m,m_s}\braket{l,S;m,m_s|l,S;j,m_j}\otimes\ket{\psi_{l,m}}\ket{S;m_s},
\end{equation}
where $\ket{\psi_{l,m}}$ is an eigenstate of the square of the orbital angular momentum and its projection on the quantization axis, $\hat{L}_z$. 
For $S=1/2$ we can write $\ket{\Phi_{j m_j}}$ in terms of the coherent spin state
\begin{equation}
	\ket{\Phi_{j m_j}}=\sum_{m} f^{j m_j}_{l m} \ket{z^{j m_j}_{l m}}\otimes\ket{\psi_{l m}}
 \label{eq:spin-state2}
\end{equation}
with $f^{j m_j}_{l m}=C^{j m_j}_{l m, \frac{1}{2} -\frac{1}{2}}
\sqrt{1+|z^{j m_j}_{l m}|^2}$, $z^{j m_j}_{l m}=C^{j m_j}_{l m, \frac{1}{2} \frac{1}{2}}/C^{j m_j}_{l m, \frac{1}{2} -\frac{1}{2}}$ and $C^{j m_j}_{l m, S m_s}=\braket{l,S;m,m_s|l,S;j,m_j}$, \eqref{eq:spin-state2} is the same as \eqref{eq:spin-state1} in the main text. Thus, the initial spin state is represented through a sum of coherent spin states.

\section{Path-Integral with Spin}
In this section we provide a more in-depth derivation of the final path-integral expression, we present the weak approximation in more detail, see \eqref{eq:RelTransSaddle} in the main manuscript, and we present a derivation of a modified saddle-point approximation that could be used for cases of strong spin-orbit coupling. We did not use the modified saddle-point approximation in the main manuscript because it is significantly more complex than the weak coupling approximation, which is valid for the case of hydrogen.
Employing the coherent state formalism, introduced above, we are in a position to utilize the path-integral formalism for a particle with spin \cite{kochetov_su_1995,pletyukhov_semiclassical_2002,pletyukhov_semiclassical_2003,morten_path_2008}, and can evaluate the kernel $\mathcal{K}_m(\pb,z)$ (defined in \eqref{eq:Kernel})
\begin{align}
 \mathcal{K}_m(\pb,z)&=\notag\\ 
        \int\displaylimits_{\rb_0} \frac{\mathcal{D}'\rb}{(2\pi)^3} &
        \int\displaylimits^{\pb_f}  \mathcal{D}'\pb
        \int\displaylimits_{z^{j m_j}_{l m}}^{z} \mathcal{D}\mu(z)
        d_{m}(\pb_0,t')
	e^{i \mathcal{A}[\rb,\pb,z,t']},
 \label{eq:K-PathIntegral}
	\intertext{where}
	\mathcal{A}[\rb,\pb,z,t']&=\notag\\
 -\int_{t'}^{\infty} d\tau&\left[
	\dot{\pb}\cdot\rb+iS\frac{z\dot{z}^*-z^* \dot{z}}{1+|z|^2}+H(\rb,\pb+\Ab(\tau),z,t')
	\right]
 \intertext{and}
	H(\rb,\pb,z,\tau)&=K(\pb)+U(\rb)+ \mathbf{C}_{SO}(\rb,\pb)\cdot\mathbf{n}(z).
\end{align}
Here, $\mathbf{n}(z)$ is a semiclassical representation of $\hat{S}$, such that $\mathbf{n}(z)=(n_1,n_2,n_3)$, $n_1+i n_2=2 z^*/(1+|z|^2)$ and $n_3=-(1-|z|^2)/(1+|z|^2)$.

\subsection{Weak-coupling limit}
\label{sec:weak-coupling}
The kernel, given by \eqref{eq:Kernel} may be solving using the saddle-point approximation, however, a simpler solution is possible via the so-called weak approximation. The first step is to use the fact that the functional integral over $z$ may be solved analytically. Rewriting \eqref{eq:K-PathIntegral} as
\begin{align}
    &\mathcal{K}_m(\pb,z)=\notag\\ 
        \int\displaylimits_{\rb_0} &\frac{\mathcal{D}'\rb}{(2\pi)^3} 
        \int\displaylimits^{\pb_f}  \mathcal{D}'\pb
        %\int\displaylimits_{z^{j m_j}_{l m}}^{z} \mathcal{D}\mu(z)
        \mathcal{M}^{m}_{\mathrm{SO}}(\rb,\pb,z,t')
        d_{m}(\pb_0,t')
	e^{i \mathcal{A}_{0,I}[\rb,\pb,t']},
 \intertext{where all spin and spin-orbit term are collected in}
    &\mathcal{M}^{m}_{\mathrm{SO}}(\rb,\pb,z,t')=
    \int\displaylimits_{z^{j m_j}_{l m}}^{z} \mathcal{D}\mu(z)
    e^{i\mathcal{A}_{\mathrm{SO}}[\rb,\pb,z,t']}.
    \label{eq:SpinOrbit-PathIntegral}
\intertext{Here,}
    &\mathcal{A}_{0,I}[\rb,\pb,t']=
    -\int_{t'}^{\infty} d\tau\left[
	\dot{\pb}\cdot\rb +H_{0,I}(\rb,\pb+\Ab(\tau),\tau)
    \right]
 \intertext{and}
 &\mathcal{A}_{\mathrm{SO}}[\rb,\pb,z,t']=
 \notag\\
    &-\int_{t'}^{\infty} d\tau\left[
        iS\frac{z\dot{z}^*-z^* \dot{z}}{1+|z|^2}+H_{SO}(\rb,\pb+\Ab(\tau),z)
    \right],
\end{align}
where $H_{0,I}(\rb,\pb,\tau)=K(\pb)+U(\rb)$ and $H_{SO}(\rb,\pb,z)=\mathbf{C}_{SO}(\rb,\pb)\cdot\mathbf{n}(z)$.
It is possible to solve \eqref{eq:SpinOrbit-PathIntegral} as \cite{pletyukhov_semiclassical_2003}
\begin{align}
    \mathcal{M}^{m}_{\mathrm{SO}}(\rb,\pb,z,t')&=
    \frac{a^*(t)-b^*(t)z^{j m_j *}_{l m}+b(t)z+a(t)z^* z^{j m_j}_{l m}}
    {\sqrt{1+|z|^2}\sqrt{1+|z^{j m_j}_{l m}|^2}},
    \label{eq:M_SO}
\end{align}
where $a(t)$ and $b(t)$ may be obtained by solving the following ordinary differential equation (ODE) \cite{pletyukhov_semiclassical_2003}
\begin{align}
\label{eq:ab-ode}
    \dot{a}&=-\frac{i}{2c^2 r}\frac{d V}{d r}\left(
	L_z(t) a - (L_x(t)-i L_y(t))b^*
	\right),\notag\\
	\dot{b}&=-\frac{i}{2c^2 r}\frac{d V}{d r}\left(
	L_z(t) b + (L_x(t)-i L_y(t))a^*
	\right)
\end{align}
with $a(0)=1$, $b(0)=0$ and $\Lb(t)=\rb(t) \times (\pb(t)+\Ab(t))$.

Now that we have a solution for $\mathcal{M}^{m}_{\mathrm{SO}}(\rb,\pb,z,t')$, we may solve the remaining path integral with the saddle-point approximation, as in the spin-less version. Here, we assume that $\mathcal{M}^{m}_{\mathrm{SO}}(\rb,\pb,z,t')$ is a slowly varying function like $d_m(p,t')$ and treat it as a prefactor in the semi-classical approximation. This is a weak-coupling approximation, such as employed in Ref.~\cite{pletyukhov_semiclassical_2003}, assuming that there is an effect of the trajectory motion on the spin but not vice versa, and will be valid if the spin-orbit action (\eqref{eq:SpinOrbit-Action} in main text) is appreciably lower than the rest of the action. Thus, the final expression for the transition amplitude is given by
\begin{align}
    &M(\pb,z)=\\
    &=-i\sum_{m,s}
    C_m(\rb_s,\pb_s,t_s)
    \mathcal{M}^{m}_{\mathrm{SO}}(\rb_s,\pb_s,z_s,t_s)
    e^{i S[\rb_s,\pb_s,t_s]}
    \label{eq:M_Full_SO}
    \intertext{with}
    &C(\rb_s,\pb_s,t_s)=\sqrt{\frac{2\pi i}{\partial^2 S/\partial t'^2}}
    \frac{e^{-i\pi\nu/2}}{\sqrt{|J|}}
    d_m(\pb_{0s},t_s),
\end{align}
which are the expressions stated in the main text.
\subsection{Modified Saddle-Point Approximation}
\label{sec:Mod-SPA}
%\textbf{I think the connection (if any) to the main text should be stated here in the beginning. What is the purpose of this section?}
In this section, we present an alternative to the weak approximation that was used in the main text to formulate the CQSFA transition amplitude that included spin-orbit coupling \eqref{eq:M_Full_SO}. The weak approximation requires that the action for spin-orbit coupling is significantly less than the remaining action. However, instead it is possible to apply the saddle-point approximation to the full path integral, which allows its use for systems with higher spin-orbit coupling, e.g., for larger atoms.
This has been called the extended phase space formulation \cite{pletyukhov_semiclassical_2003}, as it treats the real and imaginary parts of the coherent spin state coefficient like an extra component of position and momentum. 
Although this formalism has the potential to be quite powerful and descriptive, there were a number of reasons that we did not use it in the main text. Firstly, it breaks the cylindrical symmetry of the equations of motion, this means simplifications that exploit this can not be used, making the calculation more numerically intensive. Secondly, the computation of the fluctuation factor, \eqref{eq:Fluctuation} is non-trivial and goes beyond the scope of this work. Finally, for strong-field ionization of hydrogen, the weak approximation remains valid and can be very simply expressed as a prefactor term.
%We do not use this formulation in the main text but it could be valuable in settig where spin-orbit coupling is higher, e.g., larger atom.
%$S_{0,\mathrm{I}}[\rb,\pb,z,t']\approx S_{\mathrm{SO}}[\rb,\pb,z,t']>>1$. 
%\textbf{What is the meaning of the symbol $S_{0,\mathrm{I}}[\rb,\pb,z,t']$ ?}
%\textbf{I think we need to say something here adressing the controvesy with assumption in Sec. V.B, that $S_{\mathrm{SO}} << 1$?? }
Hence, to derive the modified saddle-point approximation, we evaluate \eqref{eq:K-PathIntegral1} directly via the saddle point approximation leading to
%\textbf{make the next equation fit into one column}
\begin{align}
    &M(\pb,z;\Phi_{j m_j})=\notag\\
    &\sum_{m, s} \sqrt{\frac{2\pi i}{\partial^2 S/\partial t'^2}}
    \mathcal{F}\;
    d_m(\tilde{\pb}_{0s},t_s) e^{iS[\rb_s,\pb_s,z_s,t_s]}.
\end{align}
This expression is very similar to that in \eqref{eq:SpinOrbit-PathIntegral}, except here the saddle-points are different, and the fluctuation factor $\mathcal{F}$ can be determined by
\begin{equation}
    \mathcal{F}=\int \mathcal{D}[\boldsymbol{\eta}] \exp(i \mathcal{A}^{(2)}(\boldsymbol{\eta})),
    \label{eq:Fluctuation}
\end{equation}
where $\boldsymbol{\eta}=(\delta \rb, \delta v, \delta \pb, \delta u)$ (given  $z=u-i v$), which is known as the extended phase-space vector of small variations \cite{pletyukhov_semiclassical_2002}. The second variation of the action is given by
\begin{equation}
    \mathcal{A}^{(2)}(\boldsymbol{\eta})=
    \frac{1}{2}\left( 
        \boldsymbol{\eta} \cdot\Gamma \boldsymbol{\dot{\eta}}
        -\boldsymbol{\eta} \cdot 
            \frac{\partial^2 H}{\partial\boldsymbol{\eta} \partial\boldsymbol{\eta}}
        \boldsymbol{\eta}
    \right),
\end{equation}
where $\Gamma=\begin{pmatrix}
    0 & {\id}_4\\
    -{\id}_4 & 0
\end{pmatrix}$ (known as the 8-dimensional unit symplectic matrix), with ${\id}_4$ being the 4-dimensional identity matrix. The saddle-point equations are given by
\begin{align}
	\dot{\rb}&=\frac{\partial H}{\partial \pb}
                 =\pb+\Ab(\tau)+ \mathbf{C}_{SO}(\rb,\mathbf{n}(z))\\
	\dot{\pb}&=-\frac{\partial H}{\partial \rb}
                 =-\nabla U(\rb)-\frac{1}{|\rb|}\mathbf{C}_{SO}(|\rb|\mathbf{n}(z),\pb)
\intertext{while the equations for $\dot{z}$ use $z=u-i u$}
	\dot{u}&=(1+|z|^2)^2 \frac{1}{2}\frac{\partial H}{\partial u}
 %\\
               = \mathbf{C}_{SO}(\rb,\pb)\cdot
               \begin{pmatrix}
                   1-u^2+v^2\\
                   -2uv\\
                   2u
               \end{pmatrix}
\\
	\dot{v}&=(1+|z|^2)^2 \frac{1}{2}\frac{\partial H}{\partial v}
                = \mathbf{C}_{SO}(\rb,\pb)\cdot
                \begin{pmatrix}
                    -2uv\\
                   1+u^2-v^2\\
                   2v
               \end{pmatrix},
\end{align}
which provides equations of motion for the spin. This is an alternative to what was used in the main text, where the spin-orbit coupling did not affect the equations of motion for the electron trajectories. Now the position, momentum, and spin are determined by coupled equations. 
%\textbf{Is this used in main text? What are the perspectives of these findings? I think a few thoughts should be added}
This provides a very flexible formalism to describing spin through a path integral and connect it to semiclassical equations of motion, which can provide crucial insight into the dynamics. This could also be applied to describe the spin of the residual ion, where for a multielectron description a higher dimensional representation of the coherent spin state can be used. Such as description is not restricted to spin, and a similar formalism could potentially be used to incorporate the quantum state of the light field via a coherent state path integral. 

\section{Spin Averaging}
\label{sec:Spin-Averaging}

%\textbf{In this section, we should carefully check of the the $\mathcal{P}$ notation is in line with the notation used in the main text}

We may use the coherent spin state formalism to average incoherently over initial spin orientations, for hydrogen, we use the spin state $\ket{\Phi_{1/2 \pm 1/2}}=\ket{z^{1/2 \pm1/2}_{00}}\ket{\psi_{00}(t')}$, where either $z^{1/2 \pm1/2} \rightarrow 0$ (spin down) or $z^{1/2 \pm1/2} \rightarrow \infty$ (spin up). The state $\ket{\psi_{00}(t')}$ indicates that the spatial part is initially in a $s$ state with $l=m=0$. Spatial rotations to the initial state means that all possible value of initial $z_{00}$ will be covered. We can show this explicitly by integrating over the Euler angles (see \eqref{eq:EulerIntegration1} of the main text)
\begin{align}
	\mathcal{P}_{\uparrow;}(\mathbf{p})
        %&=\int d \rho \mathcal{P}_{\uparrow}(\mathbf{p};\mathcal{R}_{\rho}\ket{\uparrow})\\
	&=\frac{1}{8\pi^2}
	\int_{0}^{2\pi}\!\!\!d\alpha
 \int_{0}^{\pi}\!\!\!d\beta
 \int_{0}^{2\pi}\!\!\!d \gamma
	\sin(\beta) \mathcal{P}_{\uparrow}(\mathbf{p}; \mathcal{R}_{\alpha \beta \gamma}\ket{\uparrow}).
\end{align}
Parameterizing the rotating as $\mathcal{R}_{\alpha \beta \gamma}=\exp(-i\alpha \sigma_z /2)\exp(-i\beta \sigma_y /2)\exp(-i\gamma \sigma_z /2)$, we may write the rotation of the spin up state as 
\begin{align}
	&\mathcal{R}_{\alpha \beta \gamma}\ket{\uparrow}=\notag\\
 &e^{i \gamma/2}(e^{-i \alpha/2}\cos(\beta/2)\ket{\uparrow} + e^{i \alpha/2}\sin(\beta)\ket{\downarrow}).
\intertext{This may be written in terms of coherent spin states}
\mathcal{R}_{\alpha \beta \gamma}\ket{\uparrow}&=e^{i\alpha/2-i\gamma/2}
\ket{e^{-i \alpha}\cot(\beta/2)}.
\end{align}		
We may drop the phasor prefactors as these will not contribute in the incoherent average, which may be written as
\begin{align}
	\mathcal{P}_{\uparrow;}(\mathbf{p})&=\notag\\
 &\hspace{-1cm}\frac{1}{8\pi^2}
	\int_{0}^{2\pi}d\alpha\int_{0}^{\pi}d\beta\int_{0}^{2\pi}d \gamma
	\sin(\beta) \mathcal{P}_{\uparrow}(\mathbf{p}; \ket{e^{-i \alpha}\cot(\beta/2)})
	\\
	&=\frac{1}{4\pi}
	\int_{0}^{2\pi}d\alpha\int_{0}^{\pi}d\beta
	\sin(\beta) \mathcal{P}_{\uparrow}(\mathbf{p}; \ket{e^{-i \alpha}\cot(\beta/2)}).
\end{align}
As there is no $\gamma$ dependence we could do the $\gamma$ integral directly. To continue, we make the variable transformation $\phi_0=-\alpha$ and $u_0=\cot(\beta/2)$. The $u_0$ integration metric can be written as $du_0=-\frac{d \beta}{2 \sin(\beta/2)^2}$, which means that $d\beta \sin(\beta)=-\frac{4 u}{(1+u^2)^2}$. This means the averaged probability may be written as
\begin{align}
	\mathcal{P}_{\uparrow;}(\mathbf{p})&=\frac{1}{\pi}\int_{0}^{2\pi} d \phi_0 \int_{0}^{\infty} u du \frac{1}{(1+u^2)^2}\mathcal{P}_{\uparrow}(\mathbf{p}; \ket{u_0 e^{i \phi_0}}).\\
	&=\frac{2}{\pi}\int_{\mathbb{C}} d^2 z_{00} \frac{1}{(1+|z_0|^2)^2} \mathcal{P}_{\uparrow}(\pb|z_0),
	\label{eq:spinAvg}
\end{align}	
where we have let $z_{00}=u_0 e^{i \phi_0}$ with $d z_{00}^2=dz_{00}\wedge dz^*_{00}=2i u du_0\wedge d\phi_0$ leading to the integral over the coherent spin states.

Now we can compute the spin average using \eqref{eq:spinAvg}. We may write the probably given an initial $z_{00}$ and final spin up using the weak-coupling formalism, but without actually applying any approximation, so it is still exact. The probability is given by
\begin{align}
	&\mathcal{P}_{\uparrow}(\pb;z_0)=\notag\\
	&\left|
	\int_{-\infty}^{\infty} \!\! d t'
	\int \mathcal{D}\rb \!\! \int \!\! \frac{\mathcal{D}' \pb}{(2\pi)^3} 
	\mathcal{M}^{\uparrow}_{\mathrm{SO}}(\rb,\pb,t';z_{0})
	d(\pb_0,t')
	e^{i S_{0,\mathrm{I}}[\rb,\pb,t']}
	\right|^2.
\end{align}
Only the term $\mathcal{M}^{\uparrow}_{\mathrm{SO}}(\rb,\pb,t',z_0)$ contains dependence on $z_0$, from \eqref{eq:M_SO} this term may be written as
\begin{align}
	\mathcal{M}^{\uparrow}_{\mathrm{SO}}(\rb,\pb,t',z_0)&=\frac{b + a z_0}{\sqrt{1+|z_0|^2}},
	\intertext{simlarly for $\mathcal{M}^{\downarrow}_{\mathrm{SO}}$ we have}
	\mathcal{M}^{\downarrow}_{\mathrm{SO}}(\rb,\pb,t',z_0)&=\frac{a - b^{*} z^*_0}{\sqrt{1+|z_0|^2}}.
\end{align}
Here we took $z_f\rightarrow\infty$ for spin up and $z_f\rightarrow 0$ for spin down. We proceed by writing in more compact notation, we collect $t'$, $\rb$ and $\pb$ into $x$ and set $\mathcal{M}_{0,I}(x)=d(x)e^{i S_{0,I}(x)}$. Now the spin-averaged probability is
\begin{widetext}
\begin{align}
	\mathcal{P}_{\uparrow}(\pb,z_0)&=
	\frac{2}{\pi}\int_{\mathbb{C}} d^2 z_0 \frac{1}{(1+|z_0|^2)^2}
	\left|
	\int \mathcal{D} x
	\frac{b(x) + a(x) z_0}{\sqrt{1+|z_0|^2}}
	\mathcal{M}_{0,I}(x)
	\right|^2\\
	&=
	\frac{2}{\pi}
	\int \mathcal{D} x \int \mathcal{D} x'
	\mathcal{M}_{0,I}(x)\mathcal{M}^{*}_{0,I}(x')
	\int_{\mathbb{C}} d^2 z_0 \frac{1}{(1+|z_0|^2)^2}
	\left(\frac{b(x) + a(x) z_0}{\sqrt{1+|z_0|^2}}\right)
	\left(\frac{b^*(x') + a^*(x') z^*_0}{\sqrt{1+|z_0|^2}}\right)\\
	&=
	\frac{2}{\pi}
	\int \mathcal{D} x \int \mathcal{D} x'
	\mathcal{M}_{0,I}(x)\mathcal{M}^{*}_{0,I}(x')
	\underbrace{
	\int_{\mathbb{C}} d^2 z_0 \frac{
	\left(b(x) + a(x) z_0\right)
	\left(b^*(x') + a^*(x') z^*_0\right)}{(1+|z_0|^2)^3}
	}_{I}	\label{eq:spinAvg_Exapnded}
\end{align}

Here we pull out the integral over spin states and do this separately
\begin{align}
	I&=\int_{\mathbb{C}} d^2 z_0 \frac{
		\left(b(x) + a(x) z_0\right)
		\left(b^*(x') + a^*(x') z^*_0\right)}{(1+|z_0|^2)^3}\\
	&=\int_{\mathbb{C}} d^2 z_0
	\frac{b(x)b^*(x')+b(x)a^*(x')z^*_0+a(x)b^*(x')z_0+a(x)a^*(x')|z_0|^2}
	{(1+|z_0|^2)^3}\\
	&=-2\int_0^{\infty} du_0\int_{0}^{2\pi} d\phi_0
	\frac{b(x)b^*(x')+b(x)a^*(x')u_0\cancel{e^{-i\phi_0}}
		+a(x)b^*(x')u_0 \cancel{e^{i\phi_0}}+a(x)a^*(x')u_0^2}{(1+u_0^2)^3}\\
	&=-\int_0^{\infty} u_0 du_0 \frac{4 \pi(b(x)b^*(x')+a(x)a^*(x')u_0^2)}{(1+u_0^2)^3}\\
	&=\pi(b(x)b^*(x')+a(x)a^*(x')).
\end{align}
\end{widetext}
Inserting the integral back into \eqref{eq:spinAvg_Exapnded} gives
\begin{align}
	\mathcal{P}_{\uparrow;}(\pb)&=
	\int \mathcal{D} x \int \mathcal{D} x'\notag\\
	&\mathcal{M}_{0,I}(x)\mathcal{M}^{*}_{0,I}(x')(b(x)b^*(x')+a(x)a^*(x'))\\
	&=\left|	\int \mathcal{D} x a(x)\mathcal{M}_{0,I}(x)\right|^2	
	+\left|\int \mathcal{D} x b(x)\mathcal{M}_{0,I}(x)\right|^2\notag\\
	&=\mathcal{P}_{\uparrow; \uparrow}(\pb)+\mathcal{P}_{\uparrow; \downarrow}(\pb).
\end{align}
Given that, $\mathcal{M}^{\uparrow}_{\mathrm{SO}}(\rb,\pb,t',z_0)\rightarrow a$ as $z_0\rightarrow\infty$ and $\mathcal{M}^{\uparrow}_{\mathrm{SO}}(\rb,\pb,t',z_0)\rightarrow b$ as $z_0\rightarrow0$. By the same logic $\mathcal{P}_{\downarrow;}(\pb)=\mathcal{P}_{\downarrow; \uparrow}(\pb)+\mathcal{P}_{\downarrow; \downarrow}(\pb)$.
%\textbf{Again, check $\mathcal{P}$ vs $P$ notation.}

\section{Characterizing the dynamical spin coefficients $a(t)$ and $b(t)$ }
\label{sec:Ana-Sol}
Given that we restrict dynamics to the $zy$-plane and the orbital angular momentum consequently is in the $x$-direction, the ODE for $a(t)$ and $b(t)$ \eqref{eq:ab-ode} may be written as
\begin{align}
    \dot{a}(t)&=-\frac{i}{2c^2 r}\frac{d V}{d r}
	L_x(t)b^*(t),\notag\\
	\dot{b}(t)&=-\frac{i}{2c^2 r}\frac{d V}{d r}
	L_x(t)a^*(t).
\end{align}
By writing $a(t)$ and $b(t)$ explicitly in terms of real and imaginary parts, this set of ODEs may be written as a 4-vector equation $\dot{\mathbf{x}}(t)=H_{\SO}(t)  \underline{\underline{\eta}} \mathbf{x}(t)$ with  $\mathbf{x}(t)=(\Re[a(t)], \Im[a(t)], \Re[b(t)], \Im[b(t)])$ and
\begin{align}
    \underline{\underline{\eta}} &=
    \begin{pmatrix}
        0& 0& 0& 1\\
        0& 0& -1& 0\\
        0& 1& 0& 0\\
        -1& 0& 0& 0&\\
    \end{pmatrix},\\
    H_{\SO}(t)&=\frac{1}{2 cr^2}\frac{d V}{dr}L_x(t)
    %\int_{\Re[t']}^{t} d\tau H_{SO}(\tau).
\end{align}
Thus $a(t)$ and $b(t)$ may be solved, see \cite{adkins_linear_2012}, as 
\begin{align}
    a(t)&=\mathbf{x}_a \cdot
    \left[
    \exp\left(\int d t H_{\SO}(t)  \underline{\underline{\eta}}\right) \mathbf{x}_0\right]\notag\\
    b(t)&=\mathbf{x}_b \cdot 
    \left[\exp\left(\int d t H_{\SO}(t) \underline{\underline{\eta}} \right) \mathbf{x}_0
    \right],
    \intertext{where the dot product with $\mathbf{x}_a(t)=(1, i, 0,0)$ selects the real and imaginary components of $a(t)$ and $\mathbf{x}_b=(0, 0, 1,i)$ selects the real and imaginary componets of $b(t)$. The initial condition is $a(0)=1$, which means $\mathbf{x}_0=(1,0,0,0)$, this leads to the solutions }
    a(t)&=\cos(S_{\SO}(t)),\notag\\
    b(t)&=-i\sin(S_{\SO}(t)),
\end{align}
which are the solutions used in the discussion in Sec.~V.B.

\section{Relativistic Force Corrections}
\label{sec:classical-corrections}
Here we consider the equations of motion with relativistic correction to the kinetic energy to better understand why the core gets less strongly probed with than without these corrections. Hamilton's equations of motion are given by
\begin{align}
\dot{\rb} = \nabla_{\pb} K(\pb),
&&
\dot{\pb} = -\nabla_{\rb} U(\rb).
\end{align}
Considering only the relativistic corrections to kinetic energy we have $K(\pb)=\frac{1}{2}\pb^2-\frac{1}{8 c^2} \pb^4$, with $\nabla_{\pb} K(\pb) =(1-\frac{1}{2 c^2}\pb^2) \pb$. Taking the derivative of $\dot{\rb}$ with respect to time and writing $\mathbf{F}=-\nabla_{\rb} U$, we obtain a Newtonian style force equation
\begin{equation}
    \ddot{\rb}=(1-\frac{1}{2c^2} \underline{\underline{M}})\mathbf{F},
\end{equation}
where $\underline{\underline{M}}$ is a matrix given by $\underline{\underline{M}}=\id_3 \pb^2 + 2\pb \otimes\pb$. Thus, the matrix $\underline{\underline{M}}$ determines by how much the force is effectively reduced by from the classical case. For example, in the case the electron is heading directly to the residual ion, $\pb=(0,0,v)$ and $\mathbf{F}=(0,0,F)$
\begin{equation}
    \underline{\underline{M}}=
    \begin{pmatrix}
    v^2 & 0 & 0\\
    0 & v^2 & 0\\
    0 & 0 & 3 v^2\\
    \end{pmatrix}
\end{equation}
leading to the equation
\begin{equation}
    \ddot{z}=(1-\frac{3 v^2}{2c^2})F,
\end{equation}
While, in the case that the electron is travelling perpendicular to force due to the residual ion, $\pb=(0,v,0)$ and $\mathbf{F}=(0,0,F)$
%\textbf{are you sure about this p? - if so the sentense above should be modified to express that the electron experiences a force perpendicular to...} and $\mathbf{F}=(0,F,0)$. 
$\underline{\underline{M}}$ is the same as before, but a different component is non-zero leading to the equation
\begin{equation}
    \ddot{z}=(1-\frac{v^2}{2c^2})F.
\end{equation}
This effective reduction of the central force will lead to the orbital radius increasing, and thus the core is probed less strongly. These aspects are discussed in Sec. VI of the main text.

\bibliography{SpinReferences}
\end{document}